\documentclass[a4paper,11pt]{article}
\pdfoutput=1 

\usepackage{jcappub} 

\usepackage[T1]{fontenc} 
\usepackage{subfigure}

\title{\boldmath Spin-mediated modulation of chaos bound violation in Lorentz-violating black hole spacetimes}

\author[a]{Chuang Yang,}
\author[a]{Deyou Chen,}
\author[b]{Bingbing Chen}

\affiliation[a]{School of Science, Xihua University, Chengdu 610039, People’s Republic of China}
\affiliation[b]{School of Mathematics, Physics and Statistics, Sichuan Minzu College, Kangding 626001, People’s Republic of China}

\emailAdd{chuangyangyc@hotmail.com}
\emailAdd{deyouchen@hotmail.com}
\emailAdd{chenbingbingedu@hotmail.com}

\abstract{We investigate how the particle spin modulates chaos bound violations in Lorentz-violating black hole spacetimes within the framework of Bumblebee gravity. Using the Mathisson-Papapetrou-Dixon formalism, we analyze the effects of spin–curvature coupling, Lorentz symmetry breaking, and the cosmological constant on the orbital instability in static spherically symmetric (A)dS black holes. We find that the particle spin provides an additional mechanism for modulating the critical angular momentum associated with the onset of violation. The Lorentz-violating parameter suppresses the violations by reducing the accessible parameter space and the critical charge. Meanwhile, a negative cosmological constant decreases the violation region but enhances the violation magnitude for allowed configurations. Our results demonstrate that the spin and Lorentz symmetry breaking jointly regulate the interplay between orbital instability and black hole horizon properties.}

\begin{document}
\maketitle
\flushbottom

\section{Introduction}
\label{Introduction}
The dynamics of black hole horizons exhibit rich chaotic behavior and extreme sensitivity to initial perturbations, making black holes natural laboratories for exploring information scrambling and nonlinear dynamics in strongly gravitating systems. Meanwhile, developments in quantum chaos have uncovered intriguing links between black hole physics and quantum information theory. An important development in this direction was achieved by Maldacena, Shenker and Stanford (MSS), who proposed a universal upper bound on the Lyapunov exponent (LE) of thermal quantum systems, known as the chaos bound \cite{MSS},

\begin{equation}
\lambda \leq \frac{2\pi k_B T}{\hbar},
\label{eq1.1}
\end{equation}

\noindent where $\lambda$ denotes the LE, $T$ represents the temperature, $k_B$ is the Boltzmann constant, and $\hbar$ is the reduced Planck constant. In the following analysis, we adopt natural units $G=c=\hbar=k_B=1$. This inequality establishes a fundamental limit on the exponential growth rate of perturbations in thermal quantum systems. Remarkably, the bound is saturated by both the Sachdev--Ye--Kitaev (SYK) model \cite{SYK1,SYK2,MS,CFCZZ} and black holes described by Einstein gravity \cite{SS1,SS2,SS3}, lending strong support to the view that black holes are among the fastest scramblers permitted by nature. This observation has motivated extensive studies of the relationship between quantum chaos, information scrambling, and holographic duality. Although departures from the chaos bound have been reported in certain quantum systems \cite{MC,AS,PZDA1}, such apparent violations can be reconciled by introducing an appropriate effective temperature \cite{RRP,JKY}, thereby preserving the validity of the chaos bound.

Similar constraints on chaotic dynamics have also been explored in classical gravitational systems. Hashimoto and Tanahashi investigated the motion of test particles near unstable trajectories in black hole spacetimes and established a close relation between the LE and the black hole surface gravity \cite{HT}. They found that the instability rate of particle trajectories satisfies

\begin{equation}
\lambda=\kappa ,
\label{eq1.2}
\end{equation}

\noindent where $\kappa$ denotes the surface gravity. This relation exhibits a remarkable similarity to the MSS chaos bound and suggests a possible connection between classical orbital instability near black holes and the exponential growth of perturbations in quantum chaotic systems. These results have motivated further studies of chaos bounds in various black hole backgrounds and gravitational theories. 

Subsequent investigations have shown that the LE can exceed the proposed upper bound in certain parameter regimes, indicating possible violations of the chaos bound in gravitational systems. Such violations are not restricted to modified gravity theories but may also occur within general relativity (GR). For example, Zhao \textit{et al.} demonstrated that, except for Reissner--Nordström (RN) and Reissner--Nordström anti-de Sitter (RN-AdS) black holes, a wide range of black hole solutions possess parameter regions where the chaos bound is violated \cite{ZLL}. Later, Lei \textit{et al.} showed that particle angular momentum can induce chaos bound violation even in RN and RN-AdS black holes \cite{LG1,LG3,YCL}. Similar behaviors have been reported for rotating charged black holes, including Kerr--Newman \cite{KG1} and Kerr--Newman-(A)dS spacetimes \cite{KG2,KG3}. Beyond these examples, chaos bound violation has been extensively investigated in various gravitational systems, including Einstein--Maxwell black holes \cite{GCYW1,LMX}, black holes coupled to nonlinear electrodynamics or anisotropic matter fields \cite{GCYW2,JLLL}, black branes \cite{LG4,DPS}, and black holes in modified gravity theories such as Einstein--Gauss--Bonnet gravity \cite{FAAC1,FAAC2} and string-inspired models \cite{KG4,KG5}. Related studies have further explored this phenomenon in diverse gravitational settings \cite{LG2,KG5,LTW1,LTW2,SPN,DG,RP,TBAZ,TBZ,HS,TM,DMM,HMTW,GT}. 

However, most previous studies have focused on the chaotic behavior of spinless particles or photons, where the particle motion is governed by geodesic equations and the chaotic properties are determined mainly by the background spacetime geometry. Although this approach has successfully clarified how black hole parameters affect the LE, it neglects the possible influence of internal particle degrees of freedom. Since elementary particles generally possess intrinsic angular momentum, the interaction between spin and spacetime curvature may provide an additional mechanism for regulating chaotic dynamics. For spinning particles, the motion in curved spacetime is described by the Mathisson--Papapetrou--Dixon (MPD) equations rather than the geodesic equations. The coupling between the particle spin tensor and the spacetime curvature introduces an additional spin--curvature interaction, which modifies the effective potential and consequently changes the properties of circular orbits and their stability. Since the LE is closely related to the instability of circular orbits through the curvature of the effective potential, particle spin can directly influence the chaotic behavior and the violation of the chaos bound. Therefore, spin provides a new dynamical degree of freedom beyond the conventional spacetime parameters for exploring black hole chaos. Meanwhile, modified gravity theories provide new theoretical frameworks for studying black hole chaos beyond Einstein gravity. Among them, Bumblebee gravity is characterized by spontaneous Lorentz symmetry breaking induced by a vector field with a nonzero vacuum expectation value \cite{VAK,LNPP}. The corresponding black hole solutions possess modified spacetime structures compared with those in GR. Since the background geometry plays a crucial role in determining orbital stability and chaotic dynamics, the Lorentz-violating effects in Bumblebee gravity may significantly modify unstable circular orbits and the associated LEs, thereby affecting the behavior of the chaos bound.

Motivated by these considerations, we investigate the chaos bound violation of spinning particles in spherically symmetric (A)dS black hole spacetimes within Bumblebee gravity. We focus on the effects of particle spin and the Lorentz-violating parameter on the LE and the corresponding chaos bound behavior. Unlike the spinless case, where chaotic dynamics are determined mainly by the background geometry, the present system incorporates two distinct modulation mechanisms: the modification of spacetime geometry and the intrinsic spin degree of freedom. Specifically, the Bumblebee parameter modifies orbital stability through changes in the black hole geometry, while the spin--curvature coupling further alters the effective potential and the properties of unstable circular orbits. The interplay between these two effects may lead to novel chaotic phenomena and shed light on the role of particle spin in regulating chaos bound violation in modified gravity backgrounds.

The remainder of this paper is organized as follows. In Sec.~\ref{sec2}, we review the spherically symmetric (A)dS black hole solutions in Bumblebee gravity and derive the equations of motion for spinning particles. In Sec.~\ref{sec3}, we analyze the combined effects of the particle spin, Lorentz-violating parameter, and the cosmological constant on chaos bound violation. Section 4 presents our conclusions and discussion.

\section{Dynamics of spinning particles in Bumblebee gravity} \label{sec2}

\subsection{Black hole solution in Bumblebee gravity}

The Bumblebee model provides a framework for realizing spontaneous Lorentz symmetry breaking by introducing a vector field $B_{\mu}$, referred to as the bumblebee field. The corresponding action is given by \cite{VAK}

\begin{eqnarray}
S&=&\int d^{4}x\sqrt{-g} \bigg [  \frac{1}{2\kappa_0}\left(R-2\Lambda\right)+\frac{\xi}{2\kappa}B^{\mu}B^{\nu}R_{\mu\nu}-\frac{1}{4} B_{\mu\nu}B^{\mu\nu}-V(B^{\mu}B_{\mu}\pm b^{2}) \bigg ] \nonumber \\
&&+\int d^{4}x\sqrt{-g}\mathcal{L}_{M},
\label{S}
\end{eqnarray}

\noindent where $\kappa_0=8\pi G/c^4$ is the gravitational coupling constant, $\Lambda$ is the cosmological constant, and $\xi$ characterizes the nonminimal coupling between the bumblebee field and gravity. The field strength of the bumblebee vector is defined by

\begin{equation}
B_{\mu\nu}=\partial_{\mu}B_\nu-\partial_{\nu}B_\mu.
\end{equation}

\noindent The matter sector is described by an electromagnetic field nonminimally coupled to the bumblebee field through the Lagrangian \cite{LNPP}

\begin{eqnarray}
\mathcal{L}_{M}
=\frac{1}{2\kappa_0}\left(F^{\mu\nu}F_{\mu\nu}
+\gamma B^{\mu}B_{\mu}F^{\alpha\beta}F_{\alpha\beta}\right),
\label{L}
\end{eqnarray}

\noindent where $F_{\mu\nu}=\partial_{\mu}A_{\nu}-\partial_{\nu}A_{\mu}$ is the electromagnetic field strength tensor associated with the four-potential $A_{\mu}$, and $\gamma$ denotes the coupling constant between the electromagnetic and bumblebee fields. Solving the corresponding field equations yields the black hole solution. The first static spherically symmetric black hole solution in Bumblebee gravity, without a cosmological constant, was obtained by Casana \textit{et al.}, who demonstrated that the Lorentz-violating parameter produces nontrivial corrections to the spacetime geometry \cite{RCPS}. Since then, a variety of exact solutions have been constructed, including charged and rotating black holes \cite{charge1,shadow2,rotation1,rotation2,rotation3,rotation4,rotation5}, wormholes \cite{wormhole}, and black hole solutions with a cosmological constant \cite{cosmological}. These developments have stimulated extensive studies of their physical properties, such as black hole thermodynamics \cite{Thermodynamics1,Thermodynamics2,Thermodynamics3}, shadows \cite{shadow1,shadow2}, quasinormal-mode spectra \cite{qn1,qn2}, gravitational lensing \cite{lensing1,lensing2}, and scattering effects \cite{SE}, considerably advancing our understanding of black hole physics in Lorentz-violating gravity.

The static spherically symmetric black hole solution with a cosmological constant is given by \cite{rotation3},

\begin{equation}
ds^{2}=	-f(r)dt^{2}	+\frac{1}{g(r)}dr^{2}+r^{2}(d\theta^{2}+	\sin^{2}\theta d\phi^{2}),
\label{m1}
\end{equation}

\noindent with

\begin{equation}
f(r) =g(r)(1+\ell)=	1-\frac{2M}{r}	+\frac{2(1+\ell)Q_{0}^{2}}{(2+\ell)r^{2}}	-\frac{1}{3}(1+\ell)\Lambda_{e}r^{2}.
\label{m2}
\end{equation}

\noindent In the above solution, $M$ and $Q_{0}$ denote the black hole mass and electric charge, respectively. Here, \(\ell=\xi b^2\) is the dimensionless Lorentz-violating parameter, where \(\xi\) denotes the nonminimal coupling between the bumblebee field and gravity, and \(b^2=b_\mu b^\mu\) characterizes the magnitude of the nonzero vacuum expectation value of the bumblebee field. The quantity $\Lambda_{e}$ is an effective cosmological constant independent of $\ell$, satisfying the relation $\Lambda=(1+\ell)\Lambda_{e}.$  Since $\Lambda$ may be either positive or negative, the solution describes both asymptotically dS and AdS black holes. Several useful limits of the solution can be obtained directly from Eq.~\eqref{m2}. In the Lorentz-invariant limit $\ell\rightarrow0$, Eq. \(\eqref{m2}\) reduces to the standard RN-(A)dS solution. When \(\Lambda_e=0\), the cosmological term vanishes, and the background spacetime reduces to the asymptotically flat charged Bumblebee black hole, in agreement with the previously obtained charged Bumblebee black hole solution \cite{LZY}. The simultaneous limit $\ell=\Lambda_{e}=0$ gives the standard RN spacetime. For the charged Bumblebee black hole considered here, the electromagnetic four-potential is given by

\begin{equation}
A_{t}=-\frac{Q_{0}}{r},
\label{m3}
\end{equation}

\noindent while the surface gravity associated with its event horizon takes the form

\begin{equation}
\kappa=\left(\frac{M}{r_h^2}-\frac{2(1+\ell)Q_0^2}{(2+\ell)r_h^3}
-\frac{1}{3}(1+\ell)\Lambda_e r_h\right)(1+\ell)^{-1/2},
\label{sf}
\end{equation}

\noindent where \(r_h\) denotes the radius of the black hole event horizon satisfying \(f(r_h)=0\). In the dS case, \(r_h\) refers to the black hole event horizon rather than the cosmological horizon. The corresponding Hawking temperature is then given by $T=\frac{\kappa}{2\pi}$.

\subsection{Dynamics of spinning particles}

The motion of a spinning test particle is no longer described by geodesic equations due to the presence of spin--curvature coupling. Instead, its dynamics are governed by the MPD equations \cite{HH1},

\begin{align}
\frac{D\tilde{p}^{\mu}}{D\tau}
&=-\frac{1}{2}R^{\mu}{}_{\nu\alpha\beta}u^{\nu}\tilde{S}^{\alpha\beta}
-\tilde{q}F^{\mu}{}_{\nu}u^{\nu},
\label{mpd1} \\
\frac{D\tilde{S}^{\mu\nu}}{D\tau}
&=\tilde{p}^{\mu}u^{\nu}-u^{\mu}\tilde{p}^{\nu},
\label{mpd2}
\end{align}

\noindent where $D/D\tau$ represents the covariant derivative along the particle trajectory, and $\tau$ denotes the affine parameter. The quantities $\tilde{p}^{\mu}$ and $u^{\mu}=dx^{\mu}/d\tau$ correspond to the four-momentum and four-velocity of the particle, respectively, while $\tilde{q}$ is the particle charge. Here, $R^{\mu}{}_{\nu\alpha\beta}$ denotes the Riemann curvature tensor, and $\tilde{S}^{\mu\nu}$ is the antisymmetric spin tensor. 

To uniquely determine the evolution of the spinning particle, an additional spin supplementary condition is required. In this work, we adopt the Tulczyjew--Dixon (TD) condition \cite{WT1,WT2},

\begin{equation}
\tilde{S}^{\mu\nu}\tilde{p}_{\nu}=0,
\label{TD}
\end{equation}

\noindent which constrains the spin tensor to be orthogonal to the particle four-momentum and removes the redundant degrees of freedom in the MPD system. Consequently, the equations of motion become closed and can be solved consistently. The particle mass $m$ and spin magnitude $\tilde{S}$ are conserved quantities associated with $\tilde{p}^{\mu}$ and $\tilde{S}^{\mu\nu}$, respectively. For convenience, we introduce the normalized variables

\begin{equation}
p^{\mu}=\frac{\tilde p^{\mu}}{m}, \qquad
S^{\mu\nu}=\frac{\tilde S^{\mu\nu}}{m}, \qquad
q=\frac{\tilde q}{m}, \qquad
S=\pm\frac{\widetilde S}{m},
\label{eq:normalized_variables1}
\end{equation}

\noindent where the sign of the spin parameter specifies the orientation of the particle spin. The positive (negative) sign corresponds to the spin aligned (anti-aligned) with the $z$-axis. In the following analysis, we follow the approach developed in Refs.~\cite{ZHA,HA,ZGWYL1,ZGWYL2,ZGWYL3} to obtain the equations of motion for spinning particles. The corresponding conserved quantities are given by

\begin{align}
S^{2} &=\frac{1}{2}S_{\mu\nu}S^{\mu\nu},\label{eq8}\\
-1&=p^{\mu}p_{\mu}. \label{eq9} 
\end{align}

We restrict our analysis to circular motion of spinning test particles on the equatorial plane of the black hole, for which $\theta=\pi/2$. Under this condition, the momentum and spin components satisfy $p^{\theta}=u^{\theta}=0$ and $S^{\theta\mu}=0$. The remaining nonvanishing components of the spin tensor are $S^{tr}$, $S^{t\phi}$, and $S^{r\phi}$. By applying the TD condition, the components $S^{tr}$ and $S^{r\phi}$ can be expressed in terms of $S^{t\phi}$ as

\begin{align}
S^{tr} &=-\frac{p_{\phi}}{p_{r}}S^{t\phi},\label{eq10}\\
S^{r\phi}&=-\frac{p_{t}}{p_{r}}S^{t\phi}.
\label{eq11}
\end{align}

\noindent Substituting Eqs.~\eqref{eq10} and \eqref{eq11} into Eq.~\eqref{eq8}, the spin invariant becomes

\begin{equation}
S^{2}=\frac{(1+\ell)r^{2}}{p_{r}^{2}}\left(S^{t\phi}\right)^{2}.
\label{eq12}
\end{equation}

\noindent This relation determines $S^{t\phi}$ in terms of the spin parameter $S$, with an undetermined overall sign. Choosing the positive branch, the nonzero components of the spin tensor are given by

\begin{align}
S^{t\phi}&=\frac{S p_{r}}{r\sqrt{1+\ell}},\label{stphi}\\
S^{tr}&=-\frac{S p_{\phi}}{r\sqrt{1+\ell}},\label{str}\\
S^{r\phi}&=-\frac{S p_{t}}{r\sqrt{1+\ell}}.
\label{srphi}
\end{align}

The spacetime possesses two Killing vector fields,
$\xi_{(t)}^{a}=(\partial/\partial t)^{a}$ and
$\xi_{(\phi)}^{a}=(\partial/\partial\phi)^{a}$,
associated with stationarity and axial symmetry, respectively. The conserved quantity corresponding to the timelike Killing vector is the particle energy $\tilde{E}$, which can be written as

\begin{eqnarray}
E=-p_t+\frac{1}{2}g_{tt}'S^{tr}-qA_t ,
\label{ee}
\end{eqnarray}

\noindent where $E=\tilde{E}/m$ represents the specific energy of the particle and the prime denotes differentiation with respect to $r$. Similarly, the axial Killing vector gives rise to the conserved total angular momentum $\tilde{J}$ along the $z$-axis,

\begin{eqnarray}
J=p_{\phi}+\frac{1}{2}g_{\phi\phi}'S^{r\phi},
\label{ej}
\end{eqnarray}

\noindent where $J=\tilde{J}/m$ is the specific angular momentum. This quantity describes the coupling between the particle spin and its orbital angular momentum. The sign of $J$ determines the orientation of the angular momentum relative to the $z$-axis, with positive (negative) values corresponding to alignment (anti-alignment). Combining Eqs.~\eqref{str}--\eqref{ej}, the temporal and azimuthal components of the particle momentum can be solved as

\begin{align}
p_t&=-\frac{E+qA_t-\dfrac{Sf'}{2r\sqrt{1+\ell}}J}
{1-\dfrac{S^2f'}{2(1+\ell)r}},
\label{pt}\\
p_\phi&=
\frac{J-\dfrac{S}{\sqrt{1+\ell}}(E+qA_t)}
{1-\dfrac{S^2f'}{2(1+\ell)r}}.
\label{pphi}
\end{align}

\noindent Substituting Eqs.~\eqref{pt} and ~\eqref{pphi} into into the momentum normalization condition~\eqref{eq9}, the radial momentum is obtained as

\begin{equation}
p_r=\pm\sqrt{\frac{1}{g}
	\left(-1+\frac{p_t^2}{f}-\frac{p_\phi^2}{r^2}\right)}.
\label{eq:pr}
\end{equation}

\noindent Here, the two signs correspond to different radial directions of motion: the positive branch describes outgoing motion, whereas the negative branch represents ingoing motion.

With the complete expressions for the particle momenta at hand, we proceed to derive the equations of motion from the MPD equations. The evolution equations for the radial and azimuthal coordinates are obtained as

\begin{align}
\frac{DS^{tr}}{Dt}
&=p^t\dot r-p^r
=\frac{S}{r\sqrt{1+\ell}}R_{\phi tt\phi}S^{t\phi}
+\frac{S}{r\sqrt{1+\ell}}R_{\phi rr\phi}S^{r\phi}\dot r,
\label{rdot}
\\
\frac{DS^{t\phi}}{Dt}
&=p^t\dot\phi-p^\phi
=-\frac{S}{r\sqrt{1+\ell}}R_{rttr}S^{tr}
-\frac{S}{r\sqrt{1+\ell}}R_{r\phi r\phi}S^{r\phi}\dot\phi
+\frac{qS}{r\sqrt{1+\ell}}A_t'.
\label{phidot}
\end{align}

\noindent Here and in what follows, an overdot denotes differentiation with respect to the coordinate time $t$. Solving Eqs.~\eqref{rdot} and \eqref{phidot} together with Eqs.~\eqref{stphi}, \eqref{str}, and \eqref{srphi}, the equations governing the radial and azimuthal motion become

\begin{align}
\dot r&=\frac{p^r}{p^t},
\label{dotr}\\
\dot\phi&=
\frac{	p^\phi\left(1-\dfrac{S^2f''}{2(1+\ell)}\right)
	+\dfrac{qS}{r\sqrt{1+\ell}}A_t'}{p^t\left(1-\dfrac{S^2f'}{2(1+\ell)r}\right)}.
\label{dotphi}
\end{align}

\noindent When $\ell=\Lambda_e=0$, the above equations reduce to the corresponding results for the RN black hole~\cite{YCL}. In the spinless limit $S=0$, the spin tensor and the spin--curvature coupling terms in the MPD equations vanish, and Eqs.~\eqref{dotr} and \eqref{dotphi} reduce to

\begin{equation}
\dot r=\frac{p^{r}}{p^{t}},
\qquad
\dot\phi=\frac{p^{\phi}}{p^{t}},
\end{equation}

\noindent with

\begin{equation}
p_{t}=-(E+qA_{t}),\qquad
p_{\phi}=J,\qquad
p_{r}=\pm\sqrt{\frac{1}{g}\left[-1+\frac{(E+qA_{t})^{2}}{f}	-\frac{J^{2}}{r^{2}}\right]}.
\end{equation}

\noindent These expressions describe the dynamics of a spinless charged particle in the same black hole background. In the further limit $S=\ell=\Lambda_e=0$, the background reduces to the RN spacetime, and the system becomes the standard spinless charged particle dynamics in the RN geometry, as considered in previous studies~\cite{LG1}.  Although both radial and azimuthal equations of motion are obtained, our subsequent analysis mainly focuses on the radial dynamics, since the effective potential governing particle motion is directly determined by the radial equation. The radial equation can therefore be written in the form of an effective potential, which will be used to identify unstable circular orbits and evaluate the LE.

\subsection{Local orbital LE}\label{sec2.3}

The effective potential provides a convenient framework for studying particle dynamics, as it encodes the combined effects of the background geometry, electromagnetic interaction, angular momentum, and spin-curvature coupling on the radial motion. The structure of the effective potential determines the permitted radial regions as well as the stability properties of particle orbits. Consequently, variations in the effective potential among different black hole spacetimes reflect the influence of the underlying spacetime geometry on particle dynamics. In~\cite{CMBWZ}, the connection between unstable circular orbits and the LE was investigated, where the exponent was shown to be related to the second derivative of the radial effective potential at the unstable circular orbit.

In the present work, the LE of a spinning test particle is evaluated by considering small radial perturbations about an unstable circular orbit located at a local maximum of the effective potential~\cite{CHL}. With respect to the coordinate time $t$, the radial equation of motion can be written as

\begin{align}
\frac{1}{2}m\dot{r}^{\,2}+\mathcal{V}_{\mathrm{eff}}=0,
\label{em}
\end{align}

\noindent where $\dot r$ is given by Eq.~\eqref{dotr}. Here, $p^{t}$ and $p^{r}$ denote the contravariant components of the four-momentum. For the metric considered here, they are related to the corresponding covariant components through $p^{t}=-\frac{p_t}{f}$
and $p^{r}=g p_r $. Here, $p_t$ and $p_r$ are given by Eqs.~\eqref{pt} and~\eqref{eq:pr}, respectively. Accordingly, the effective potential employed in the numerical analysis is defined as

\begin{equation}
\mathcal{V}_{\mathrm{eff}}
=
-\frac{m}{2}
\left(\frac{p^{r}}{p^{t}}\right)^2.
\label{veff}
\end{equation}

\noindent For fixed background parameters and particle parameters $(q,J,S)$, the conserved energy $E$ is not prescribed independently. Instead, $E$ and the circular orbit radius $r_0$ are determined simultaneously for each parameter set from

\begin{align}
\mathcal{V}_{\mathrm{eff}}(r_0;E)&=0,
\label{circular1}\\
\left.
\frac{\partial \mathcal{V}_{\mathrm{eff}}(r;E)}
{\partial r}
\right|_{r=r_0}&=0.
\label{circular2}
\end{align}

\noindent Here, the dependence of $\mathcal{V}_{\mathrm{eff}}$ on the fixed background and particle parameters is left implicit for notational simplicity. Equations~\eqref{circular1} and~\eqref{circular2} are solved simultaneously for the pair $(E,r_0)$. Since the present analysis focuses on unstable circular orbits, the resulting solutions are further required to satisfy

\begin{equation}
\left.
\frac{\partial^{2}\mathcal{V}_{\mathrm{eff}}(r;E)}
{\partial r^{2}}
\right|_{r=r_0}<0,
\label{unstablecondition}
\end{equation}

\noindent which identifies $r_0$ as a local maximum of the radial effective potential and hence as a radially unstable circular orbit.

To determine the local instability of such an orbit, we consider a small radial perturbation around $r=r_0$,

\begin{equation}
r(t)=r_0+\epsilon(t),
\end{equation}

\noindent where $\epsilon(t)$ denotes a small deviation from the circular orbit. Expanding Eq.~\eqref{em} around $r_0$ and using the circular orbit conditions $\mathcal{V}_{\mathrm{eff}}(r_0)=0$ and $\mathcal{V}_{\mathrm{eff}}'(r_0)=0$, we obtain

\begin{equation}
\frac{1}{2}
\left[
m\dot{\epsilon}^{\,2}
+\mathcal{V}_{\mathrm{eff}}^{\prime\prime}(r_0)\epsilon^{2}
\right]
+\mathcal{O}(\epsilon^{3})
\simeq0.
\label{ex}
\end{equation}

\noindent Retaining the leading-order contribution, the radial perturbation grows as $\epsilon(t)=\epsilon(0)e^{\lambda t}$, and hence

\begin{equation}
r(t)=r_0+\epsilon(0)e^{\lambda t}.
\end{equation}

\noindent The corresponding growth rate $\lambda$ is the local orbital LE and is determined by the curvature of the effective potential at the unstable circular orbit,

\begin{align}
\lambda^2 =
-\frac{\mathcal{V}_{\mathrm{eff}}^{\prime\prime}(r_0)}{m}=
\left.
\frac{1}{2}\frac{d^2}{dr^2}
\left(\frac{p^{r}}{p^{t}}\right)^2
\right|_{r=r_0}.
\label{le}
\end{align}

\noindent A real and positive value of $\lambda$ indicates the exponential growth of a small radial perturbation and therefore characterizes the local instability of an unstable circular orbit. Since $\lambda$ is obtained by linearizing the radial motion around a given circular orbit, it should be regarded as a local orbital instability exponent rather than as a diagnostic of global chaos in the full phase space.

\section{Numerical test of the chaos bound} \label{sec3}

\begin{figure*}[t]
	\centering
	
	\subfigure[]{
		\includegraphics[width=0.32\textwidth]{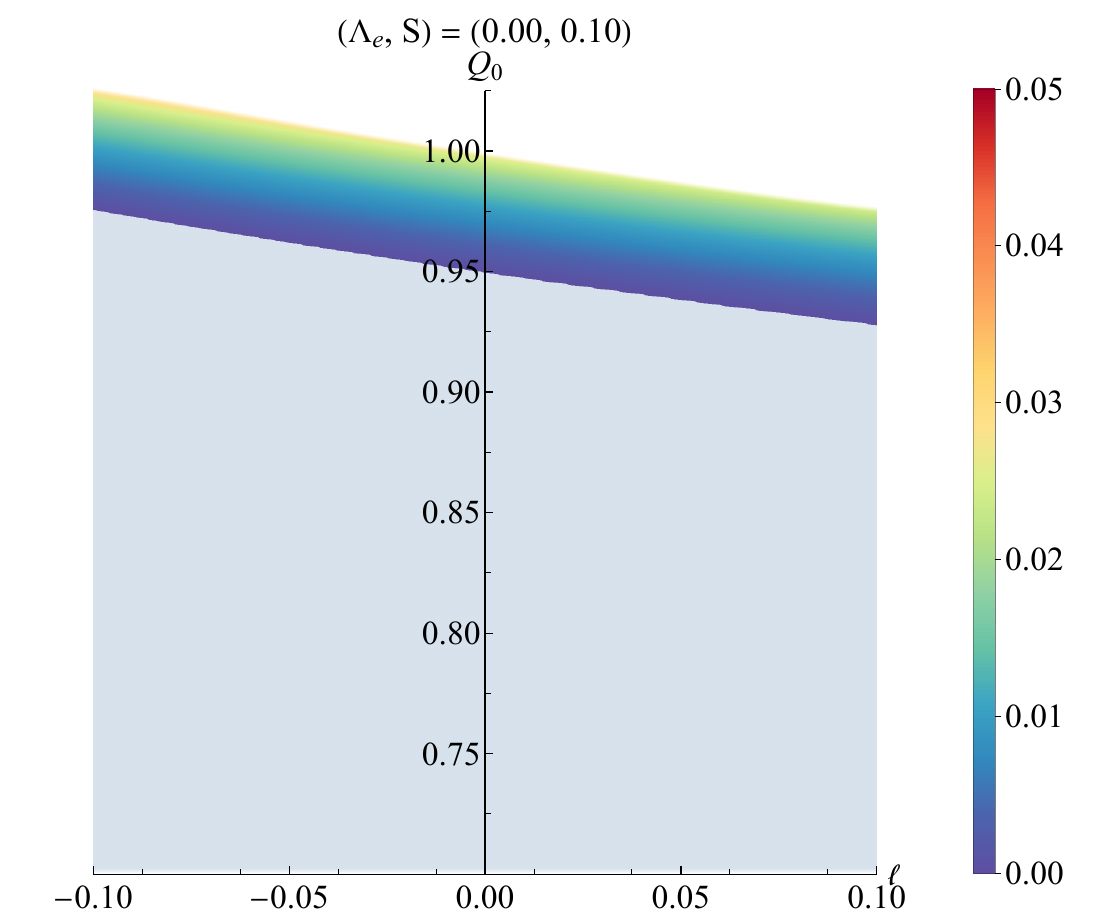}}
	\subfigure[]{
		\includegraphics[width=0.32\textwidth]{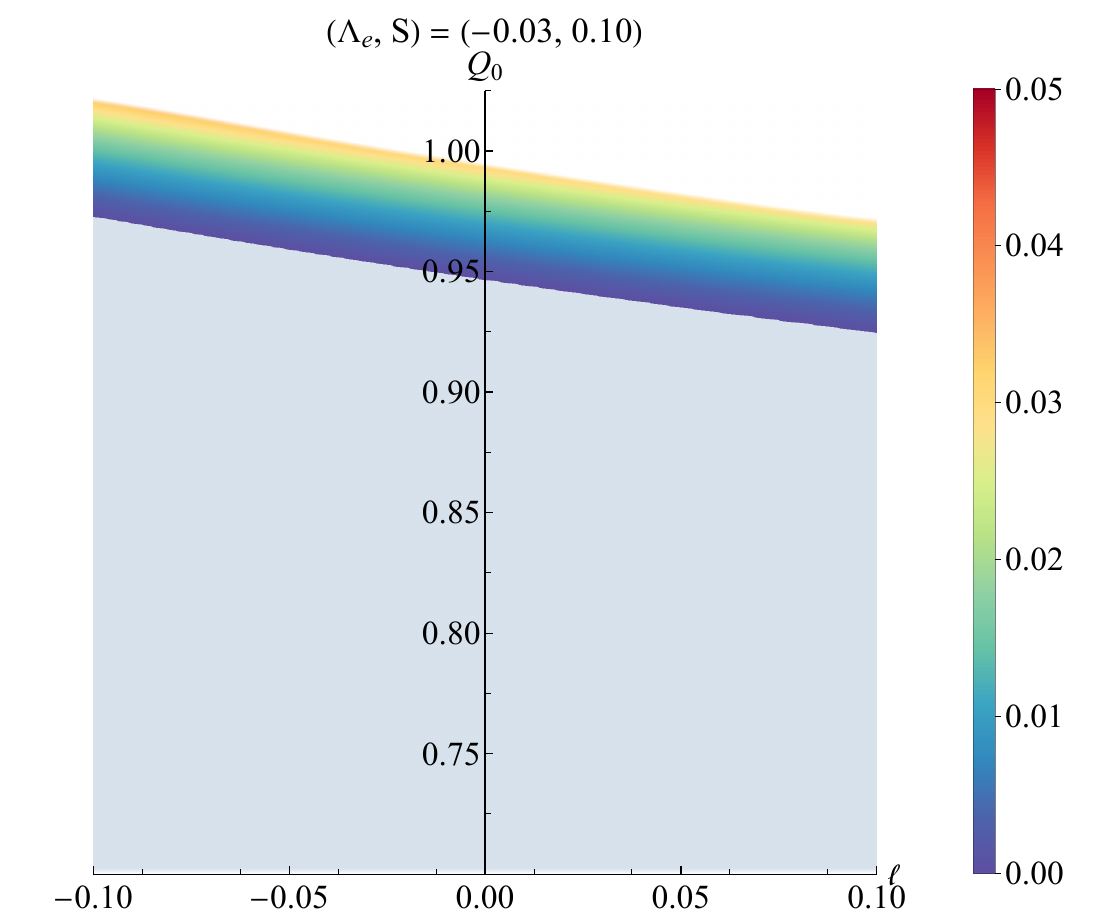}}
	\subfigure[]{
		\includegraphics[width=0.32\textwidth]{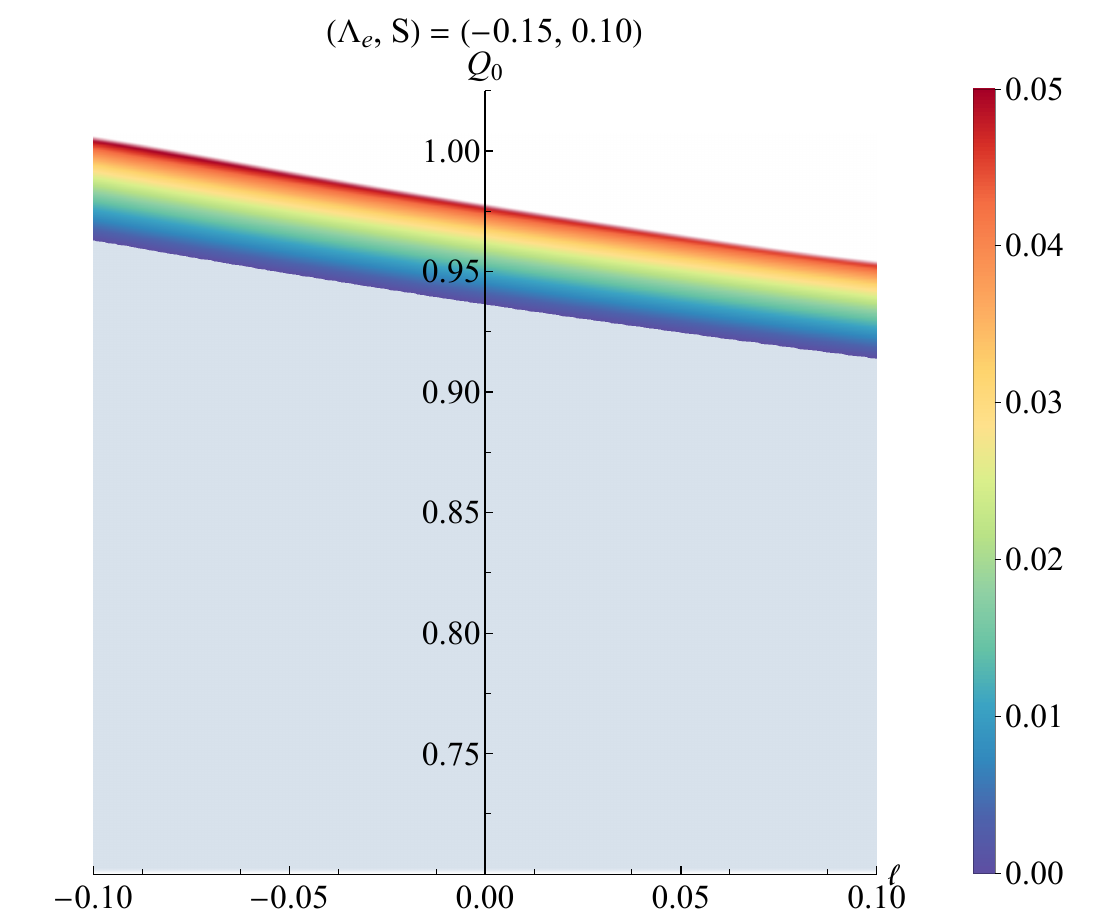}}
\caption{Effects of the black hole charge and the Lorentz-violating parameter on the chaos bound violation for $J=10.00$.}
\label{Fig1}
\end{figure*}

\begin{figure*}[t]
	\centering
	
	\subfigure[]{
		\includegraphics[width=0.32\textwidth]{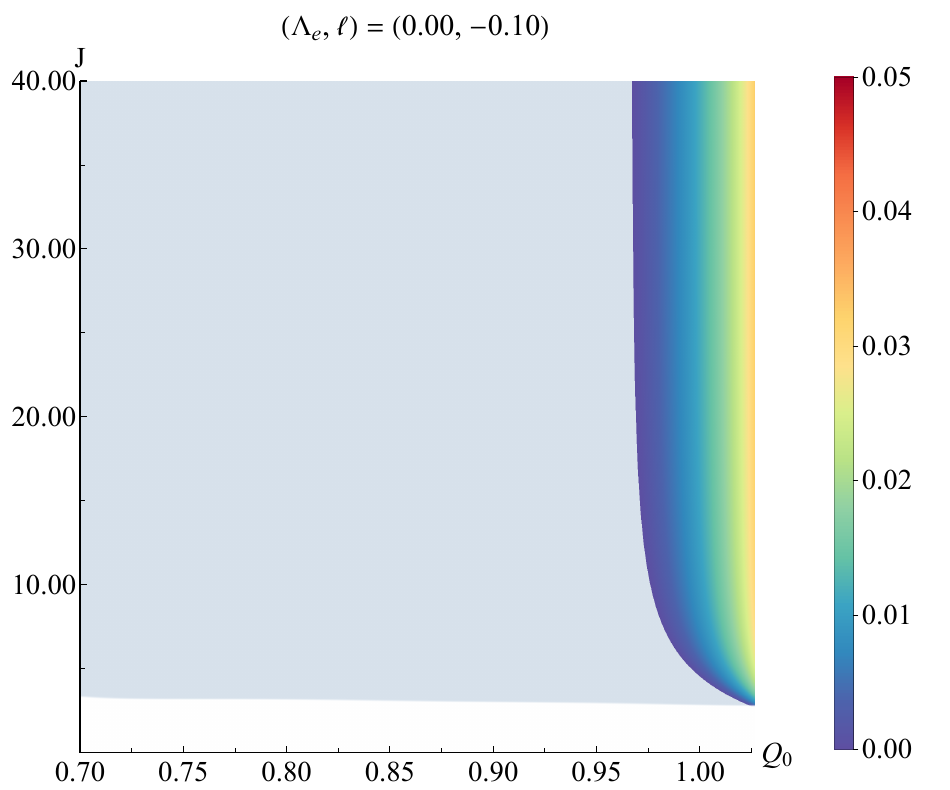}}
	\subfigure[]{
		\includegraphics[width=0.32\textwidth]{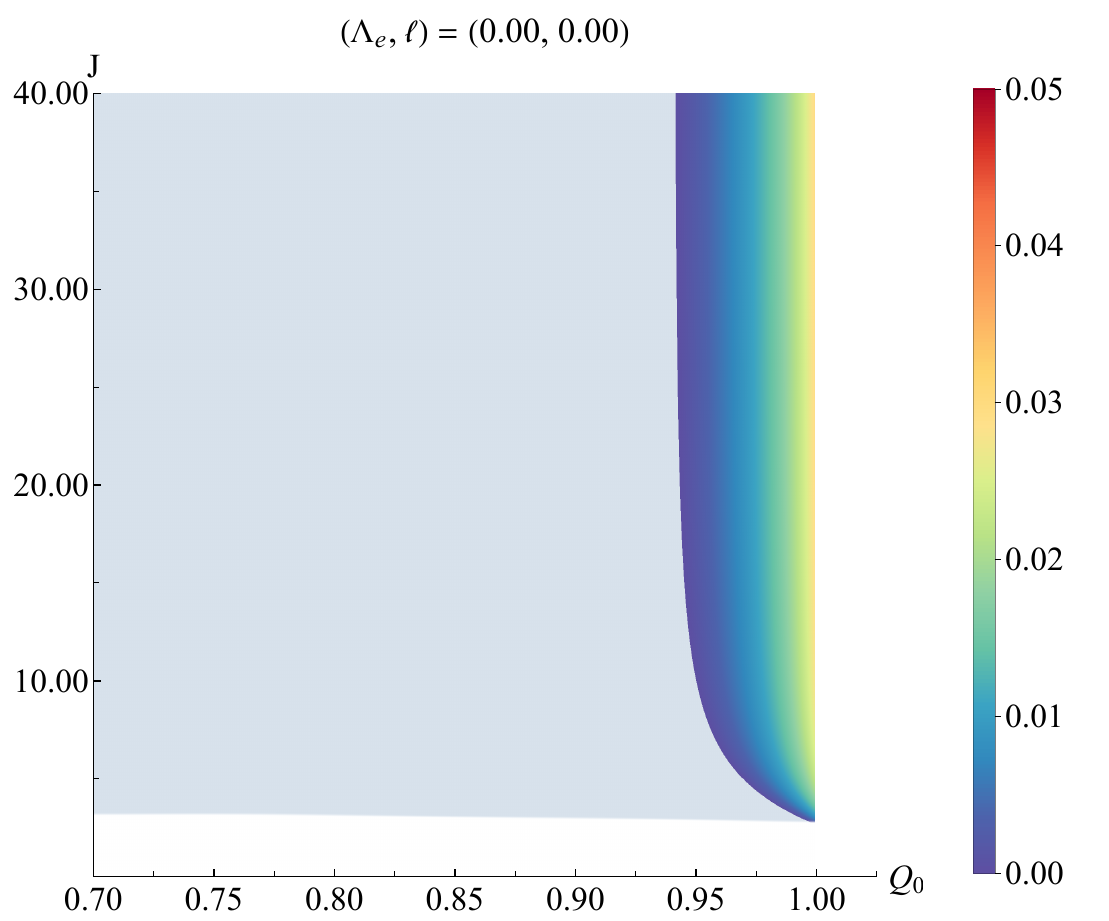}}
	\subfigure[]{
		\includegraphics[width=0.32\textwidth]{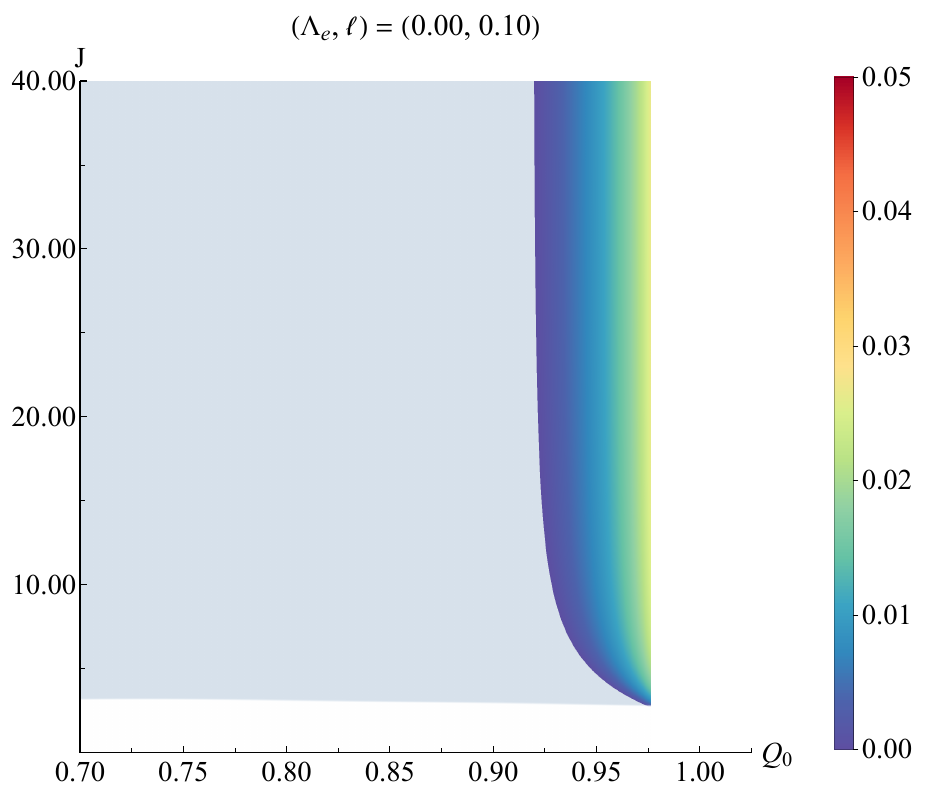}}
	
	\vspace{5mm}
	
	\subfigure[]{
		\includegraphics[width=0.32\textwidth]{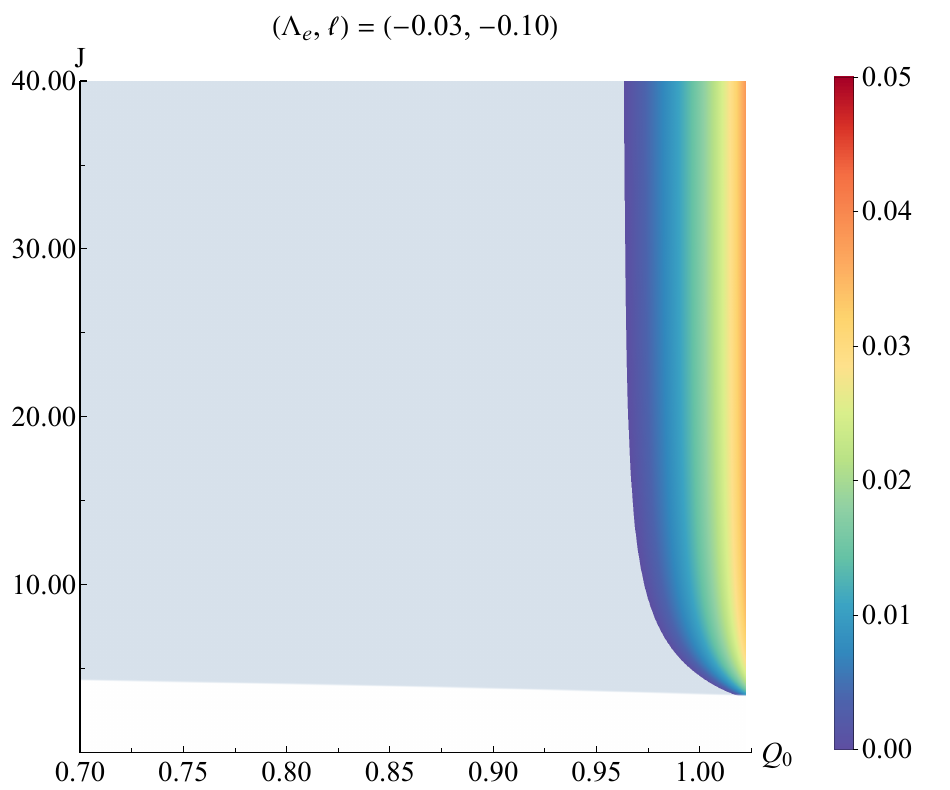}}
	\subfigure[]{
		\includegraphics[width=0.32\textwidth]{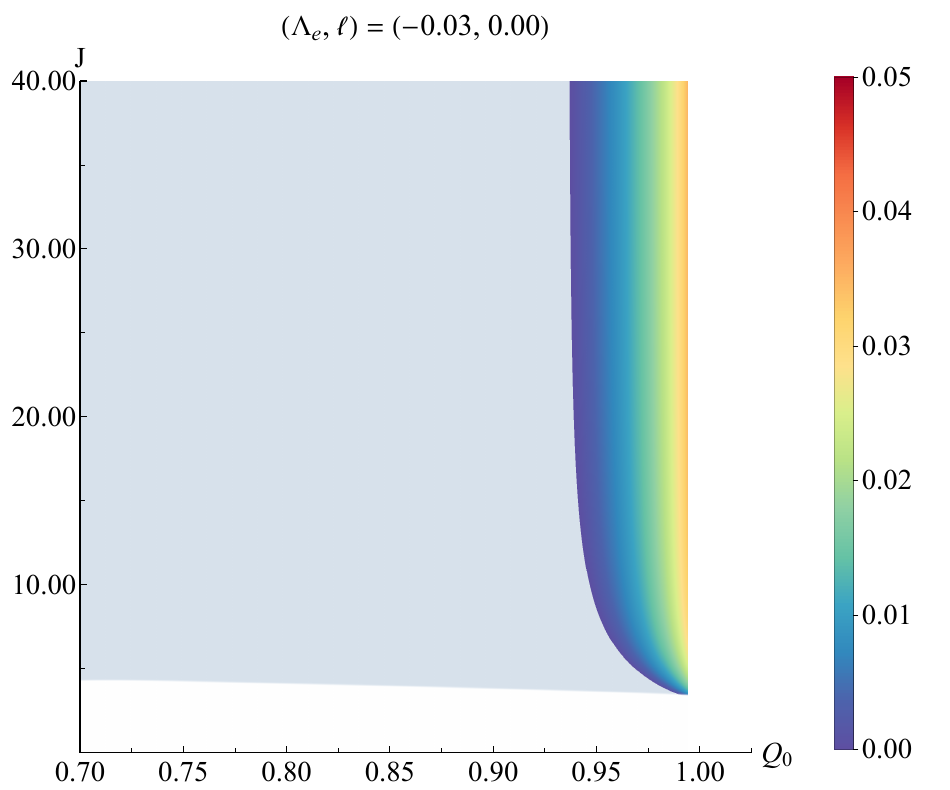}}
	\subfigure[]{
		\includegraphics[width=0.32\textwidth]{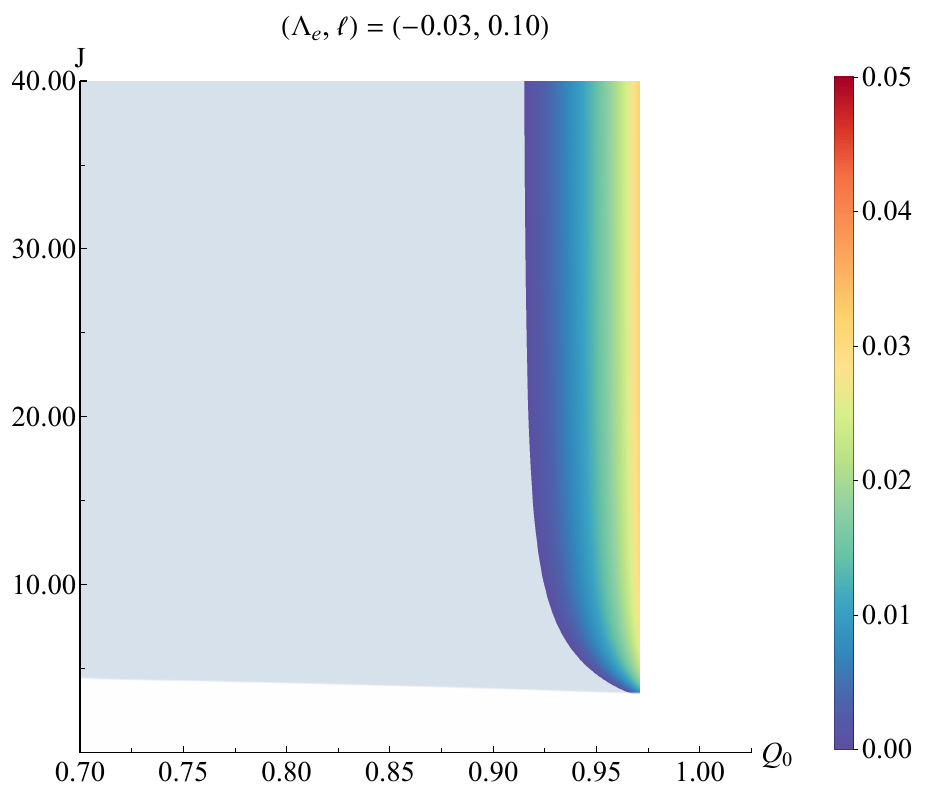}}	
	
	\vspace{5mm}
	
\subfigure[]{
\includegraphics[width=0.32\textwidth]{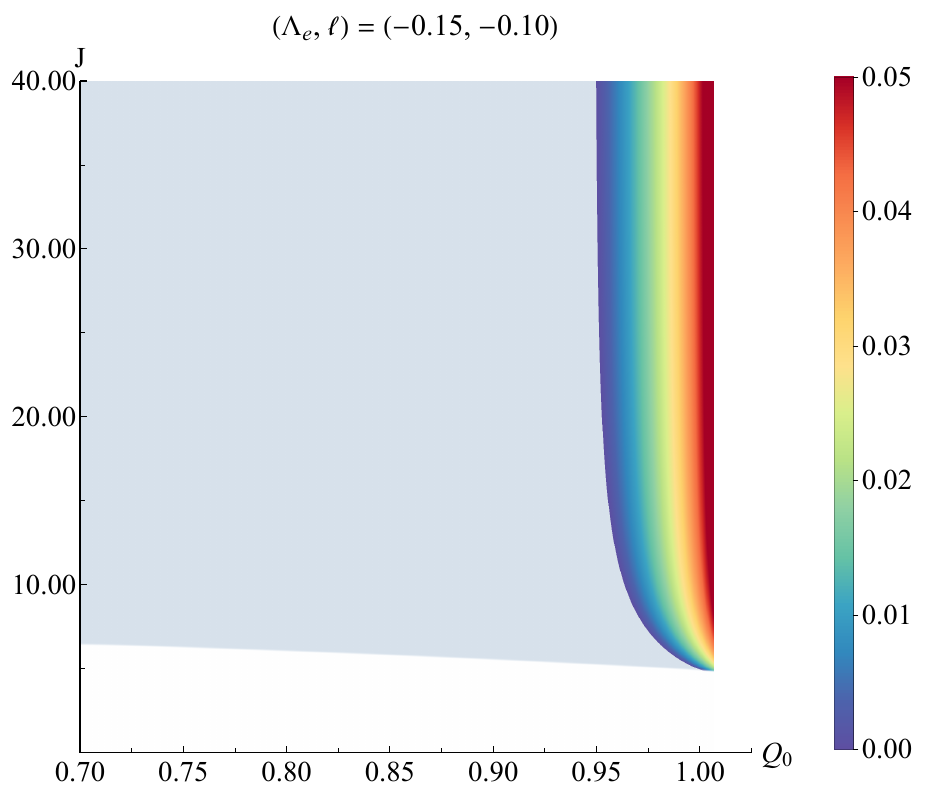}}
\subfigure[]{
\includegraphics[width=0.32\textwidth]{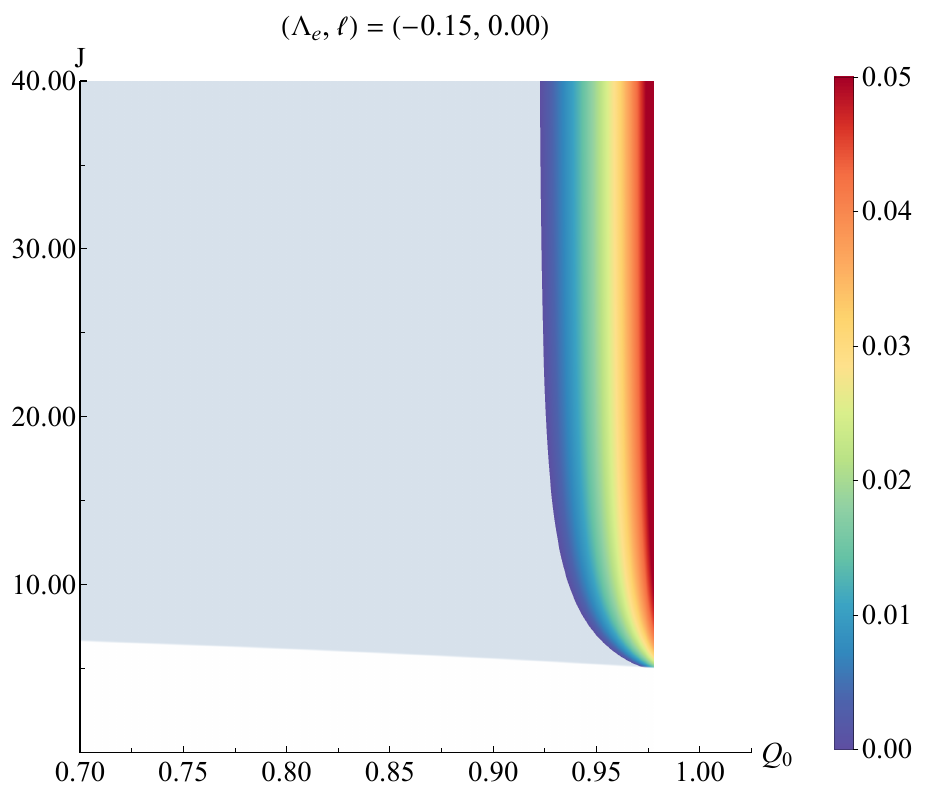}}
\subfigure[]{
\includegraphics[width=0.32\textwidth]{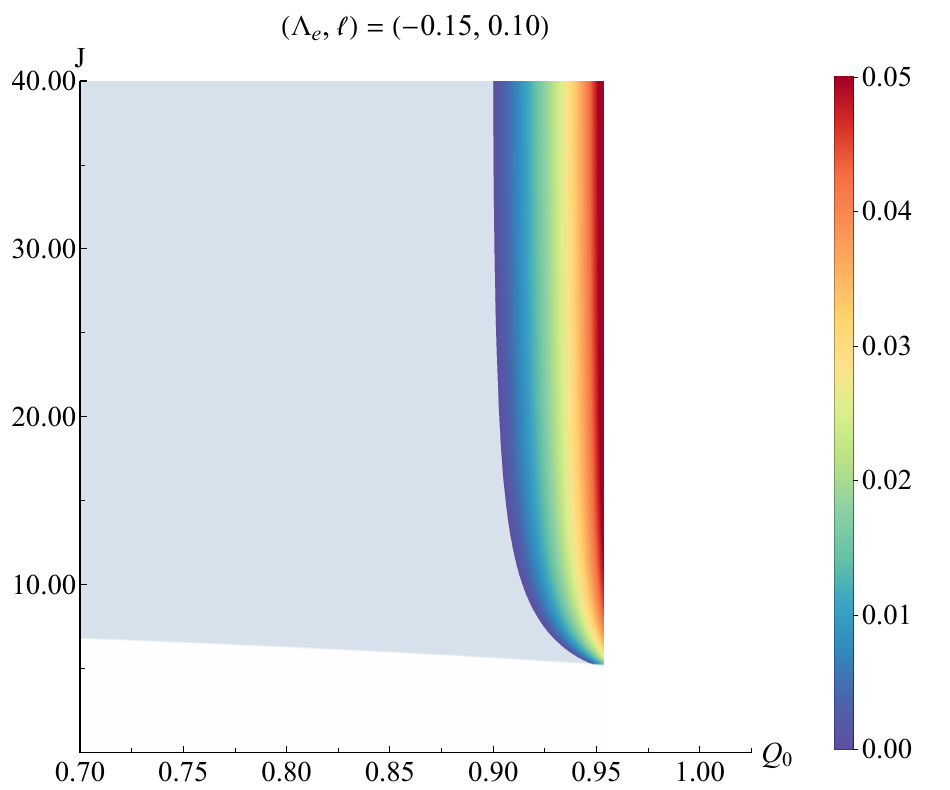}}
	
	\caption{Effects of the black hole charge and the particle total angular momentum on the chaos bound violation for $S=0.10$.}
	\label{Fig2}
\end{figure*}

\begin{figure*}[t]
	\centering
	
	\subfigure[]{
		\includegraphics[width=0.32\textwidth]{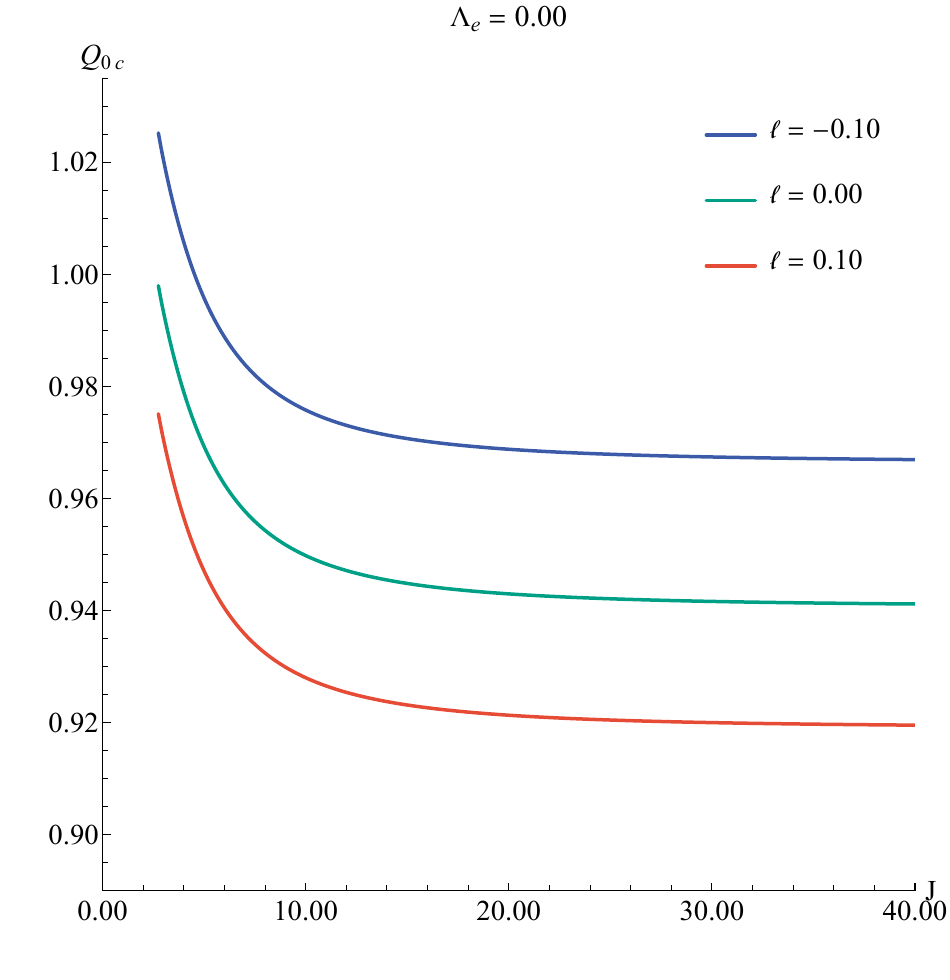}}
	\subfigure[]{
		\includegraphics[width=0.32\textwidth]{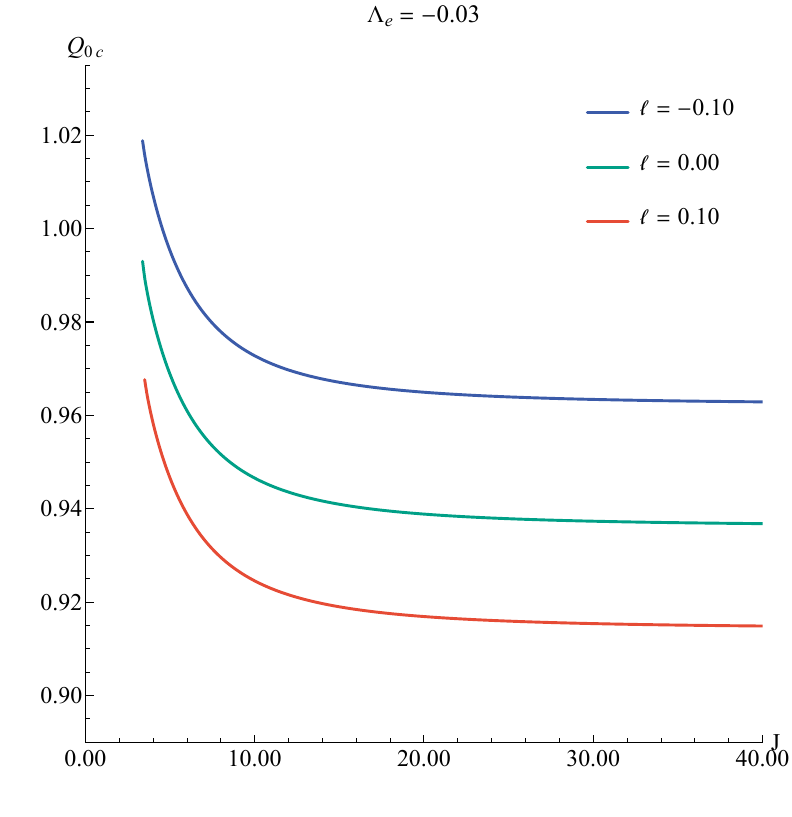}}
	\subfigure[]{
		\includegraphics[width=0.32\textwidth]{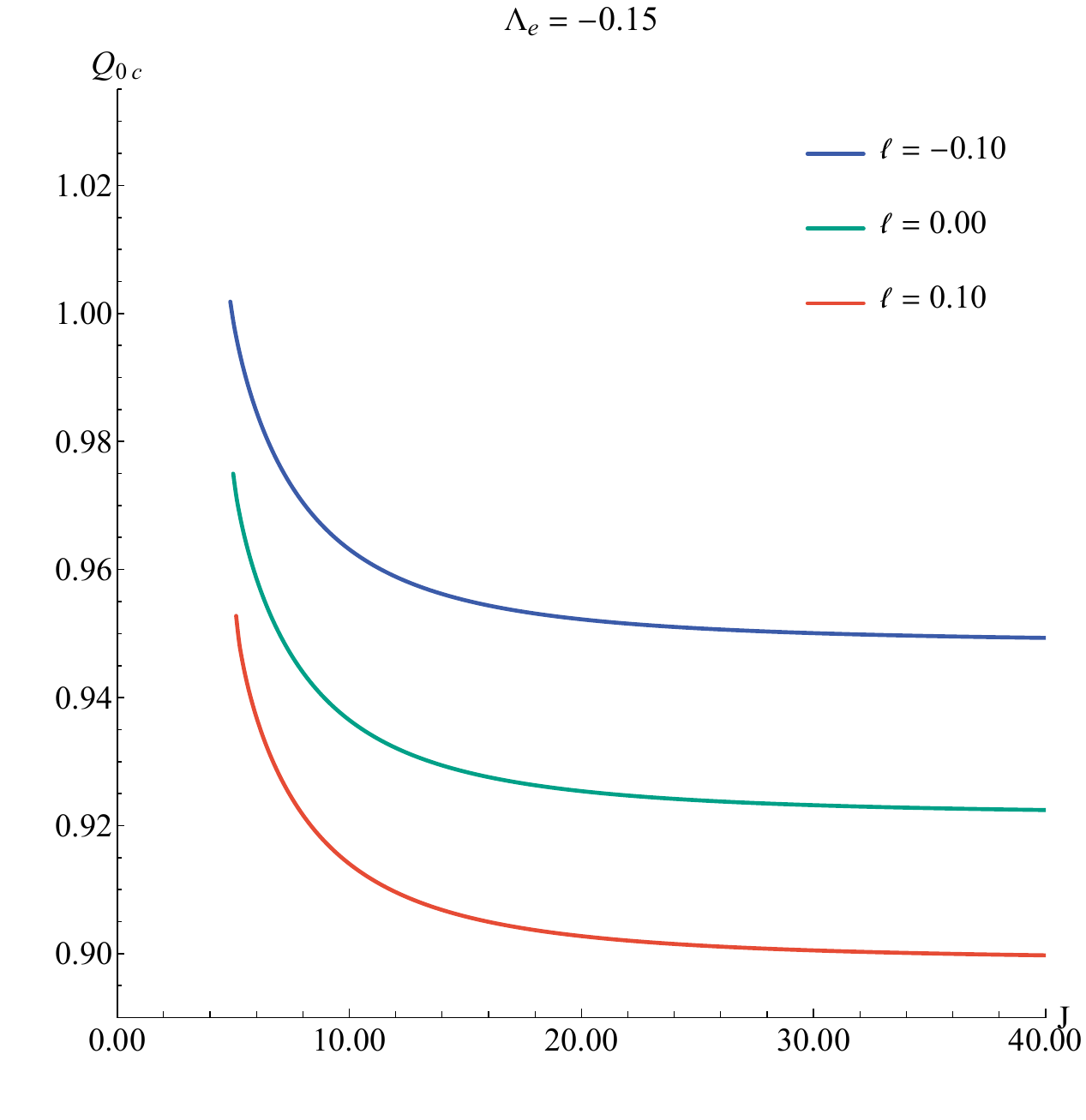}}
	
	\vspace{5mm}
	
	\subfigure[]{
		\includegraphics[width=0.32\textwidth]{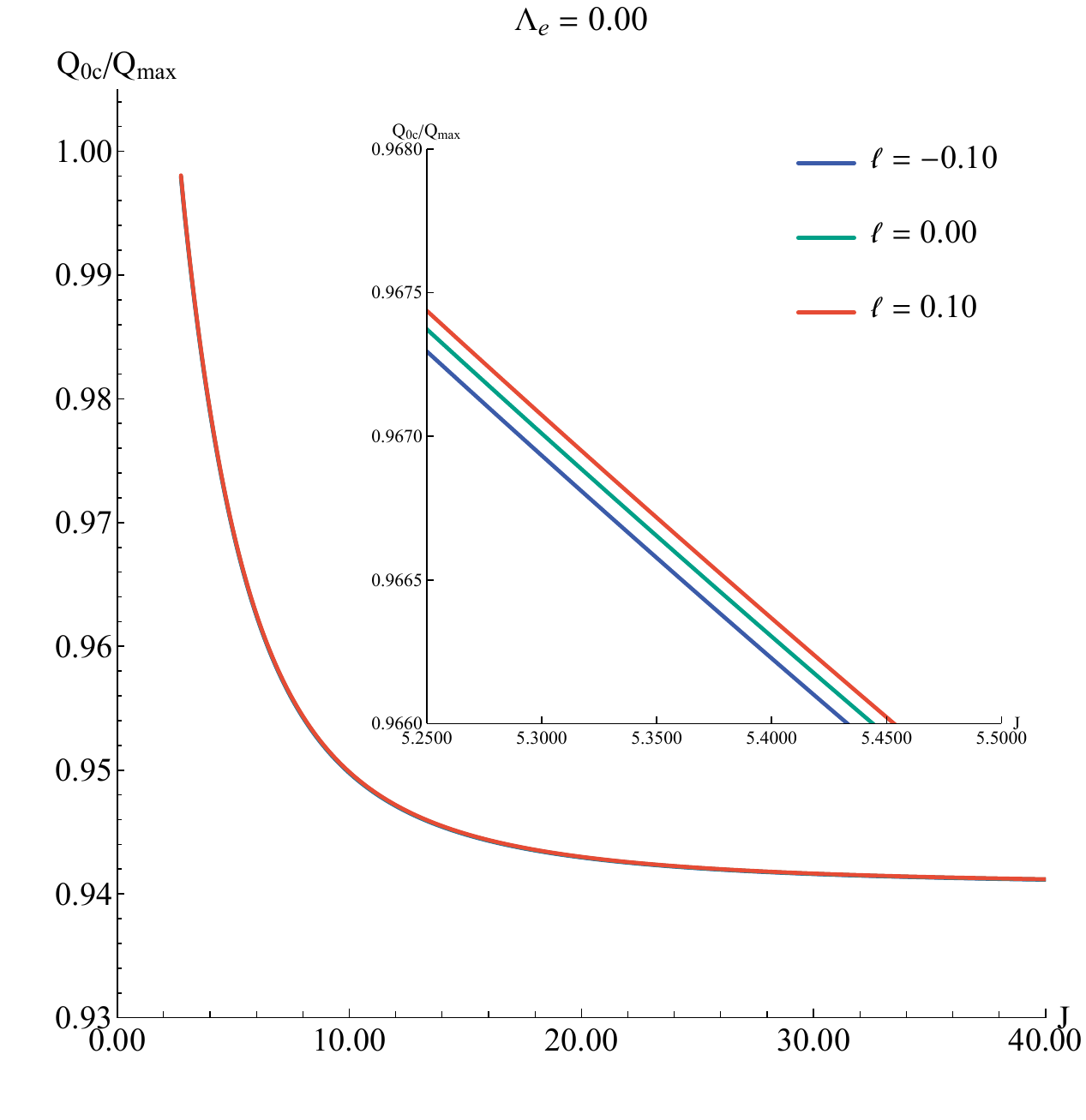}}
	\subfigure[]{
		\includegraphics[width=0.32\textwidth]{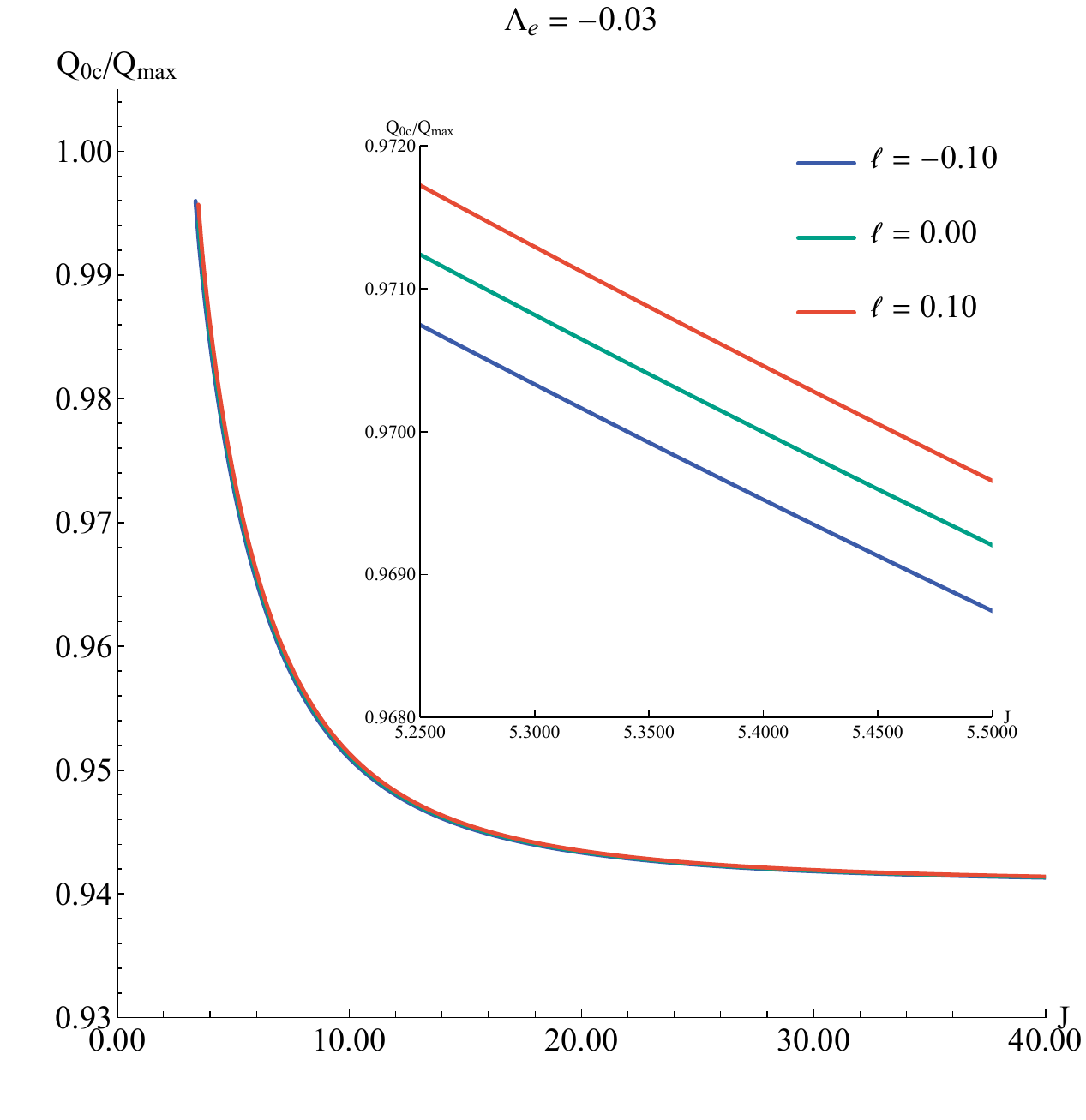}}
	\subfigure[]{
		\includegraphics[width=0.32\textwidth]{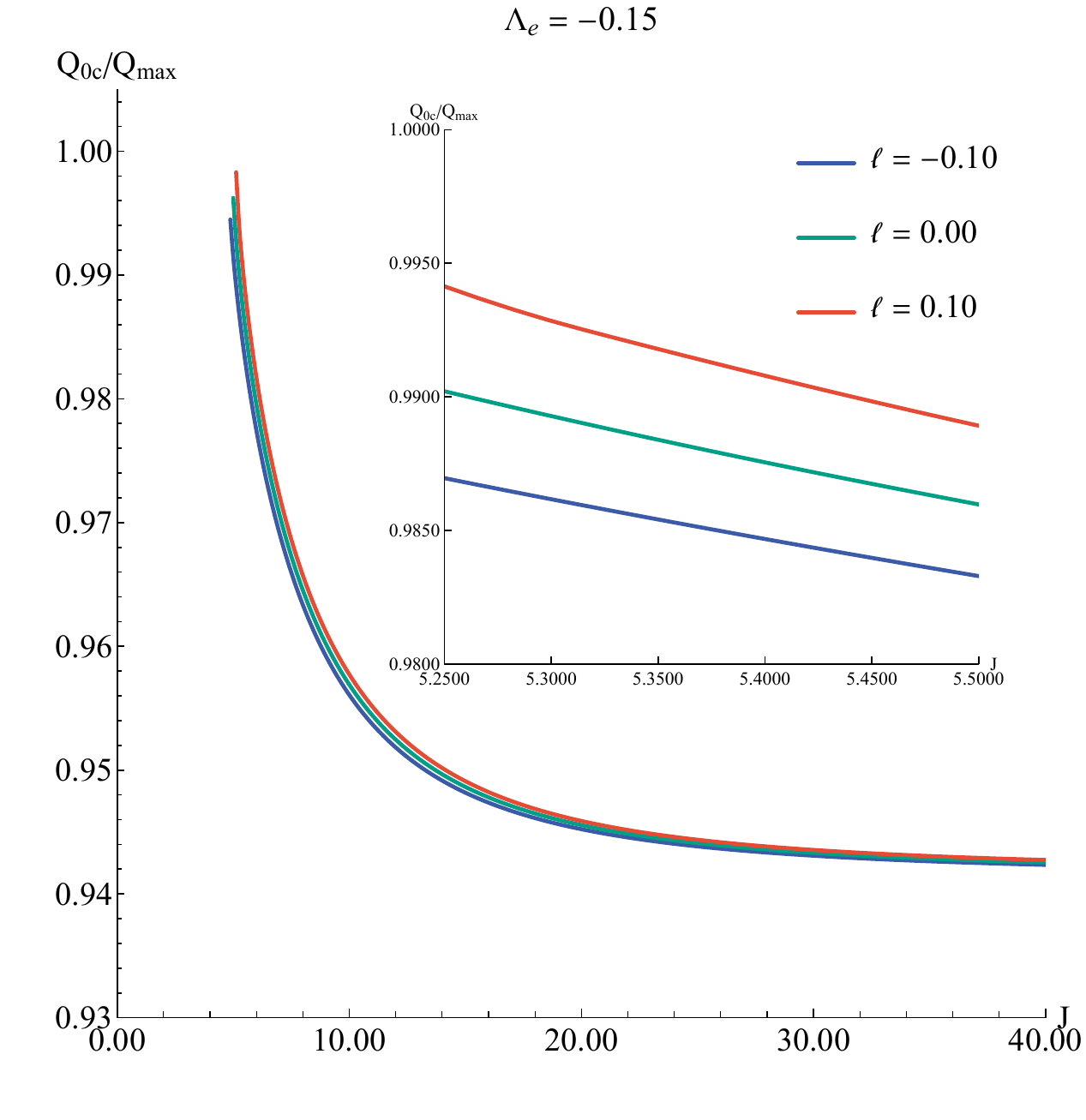}}
	
	\caption{The critical charge as a function of the total angular momentum of the particle for $S=0.10$.}
	\label{Fig3}
\end{figure*}

\begin{figure*}[t]
	\centering
	
	\subfigure[]{
		\includegraphics[width=0.32\textwidth]{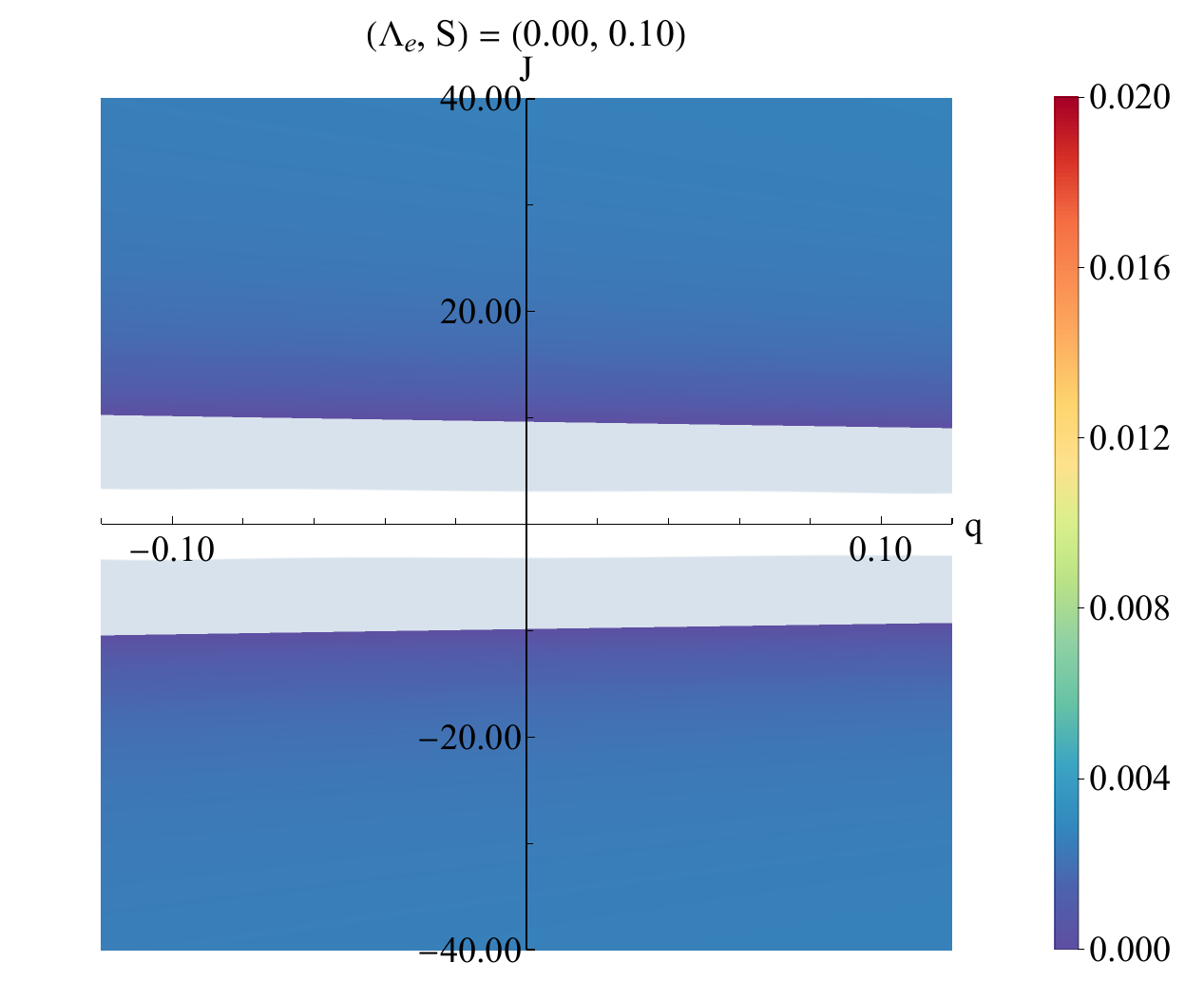}}
	\subfigure[]{
		\includegraphics[width=0.32\textwidth]{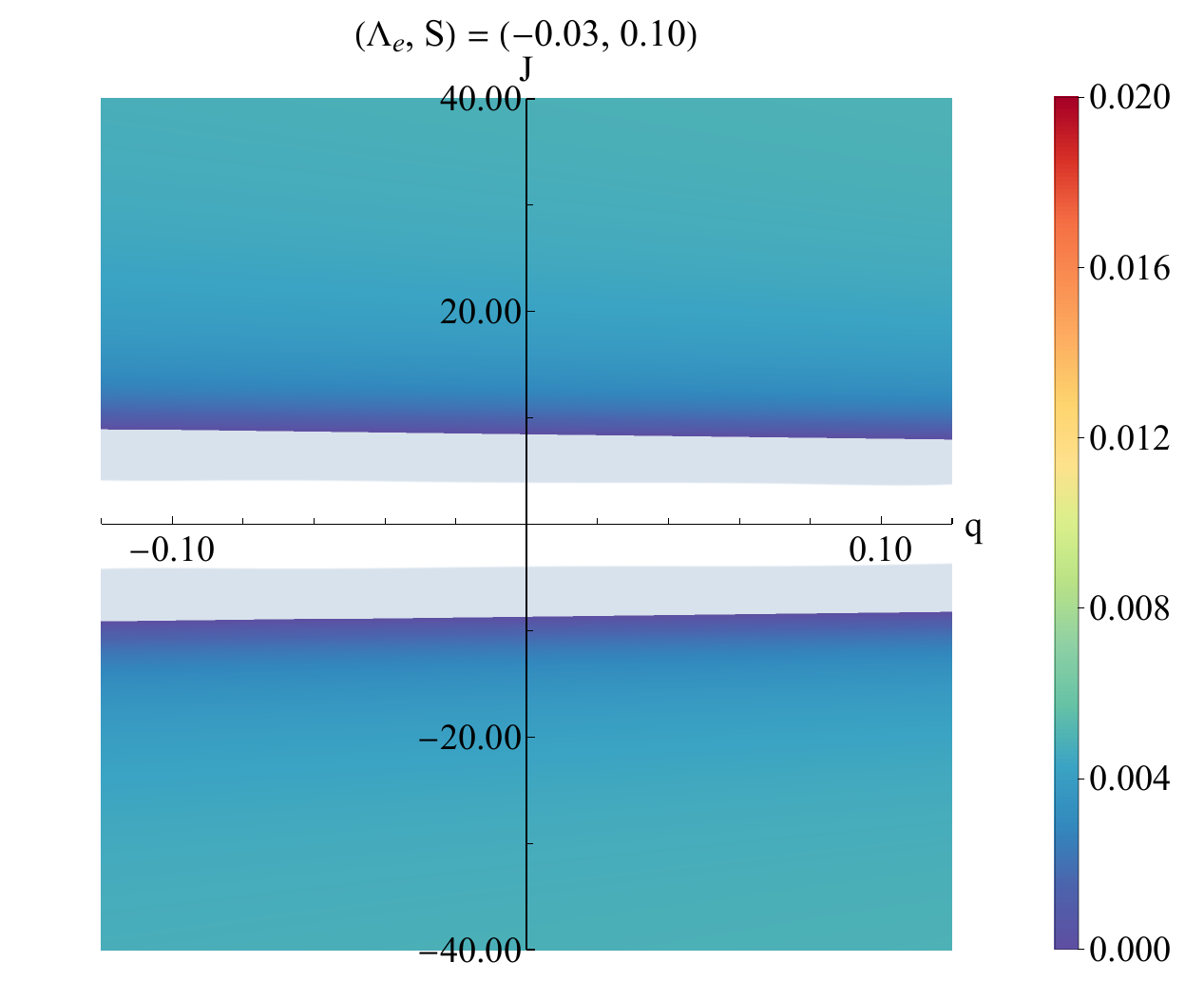}}
	\subfigure[]{
		\includegraphics[width=0.32\textwidth]{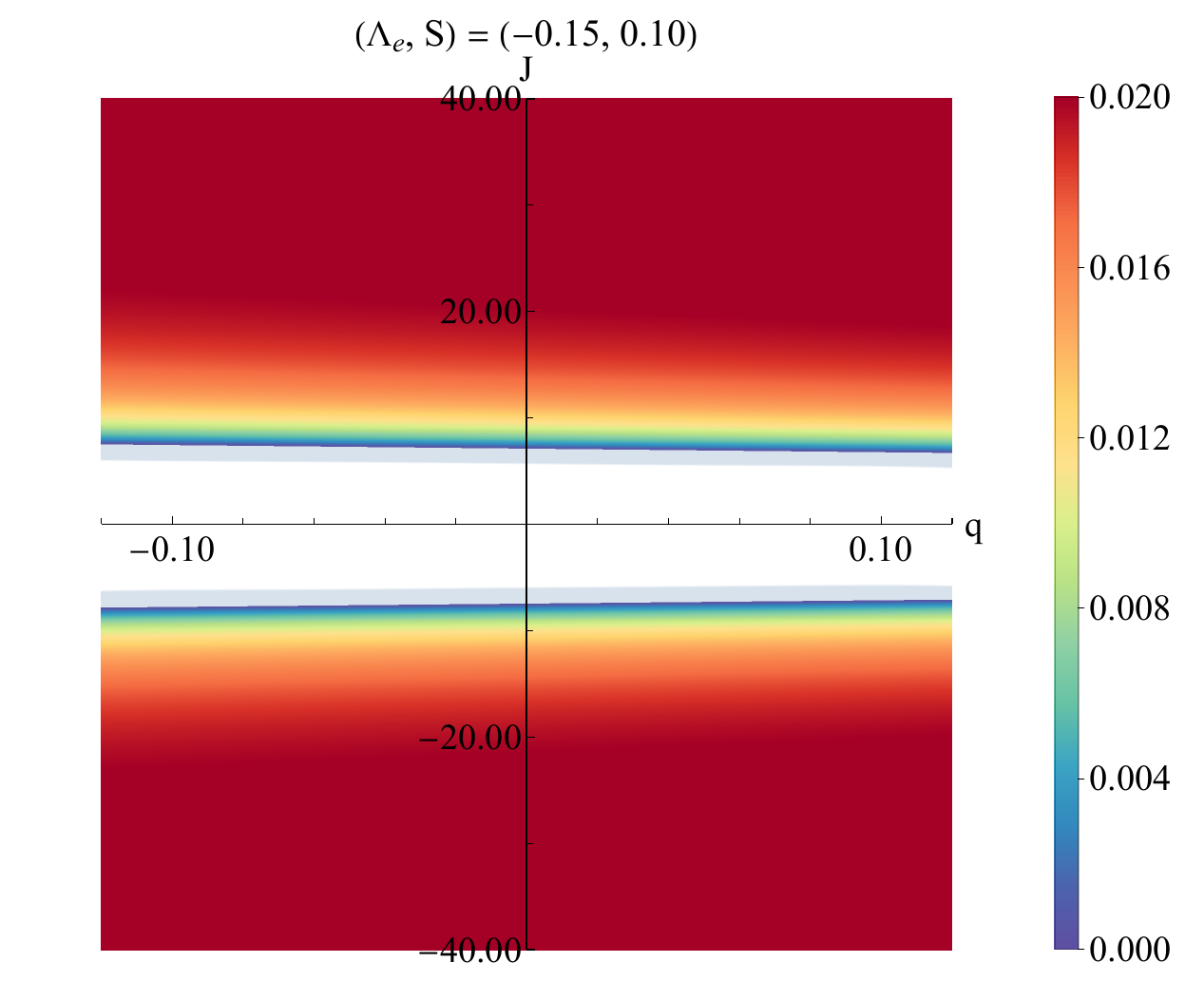}}
		\caption{Effects of the particle total angular momentum and charge on the chaos bound violation for $Q_0=0.94$, $\ell=0.05$ and $S=0.10$.}
	\label{Fig4}
\end{figure*}

\begin{figure*}[t]
	\centering
	\subfigure[]{\includegraphics[width=0.4\textwidth]{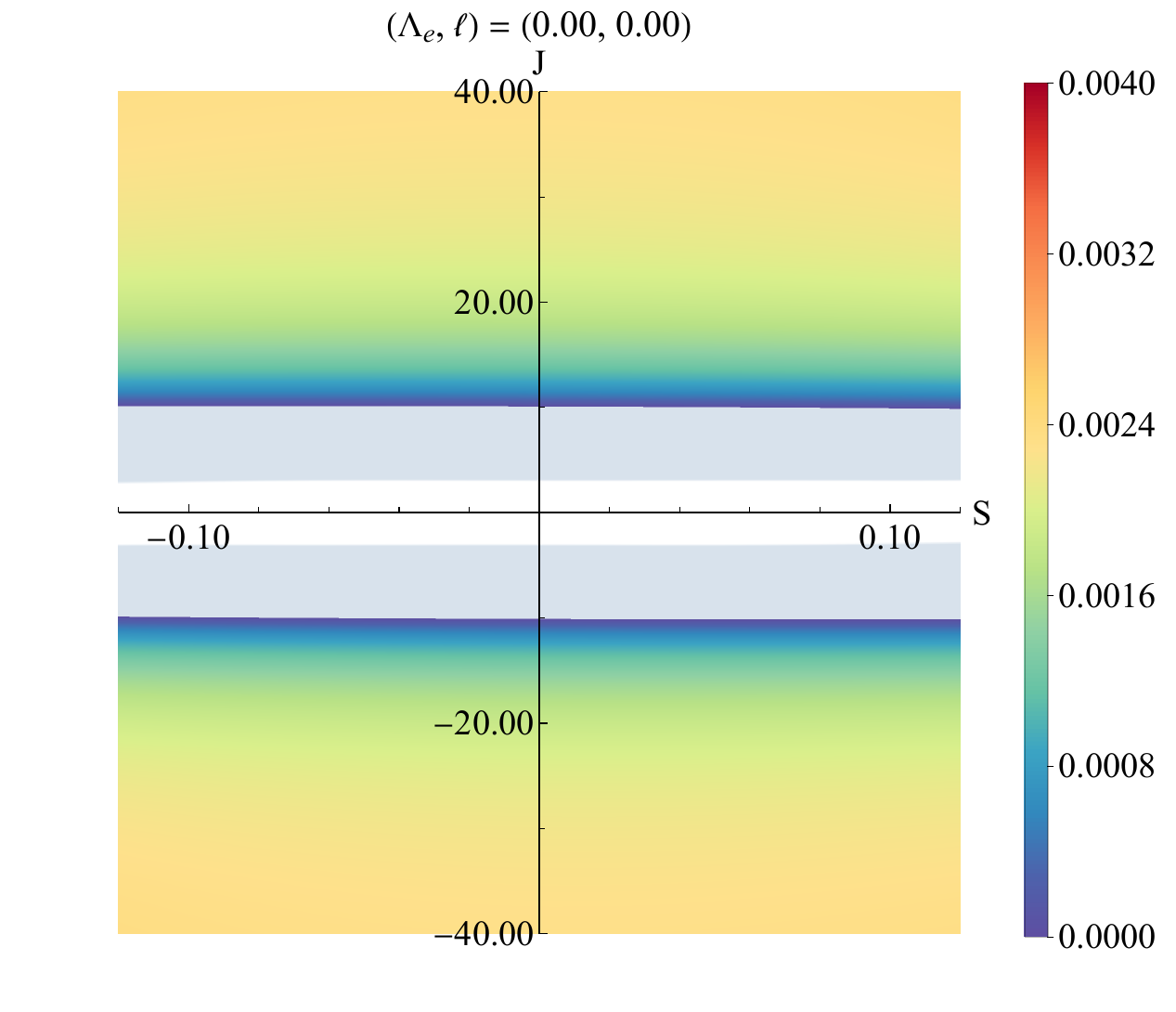}}
	\subfigure[]{\includegraphics[width=0.4\textwidth]{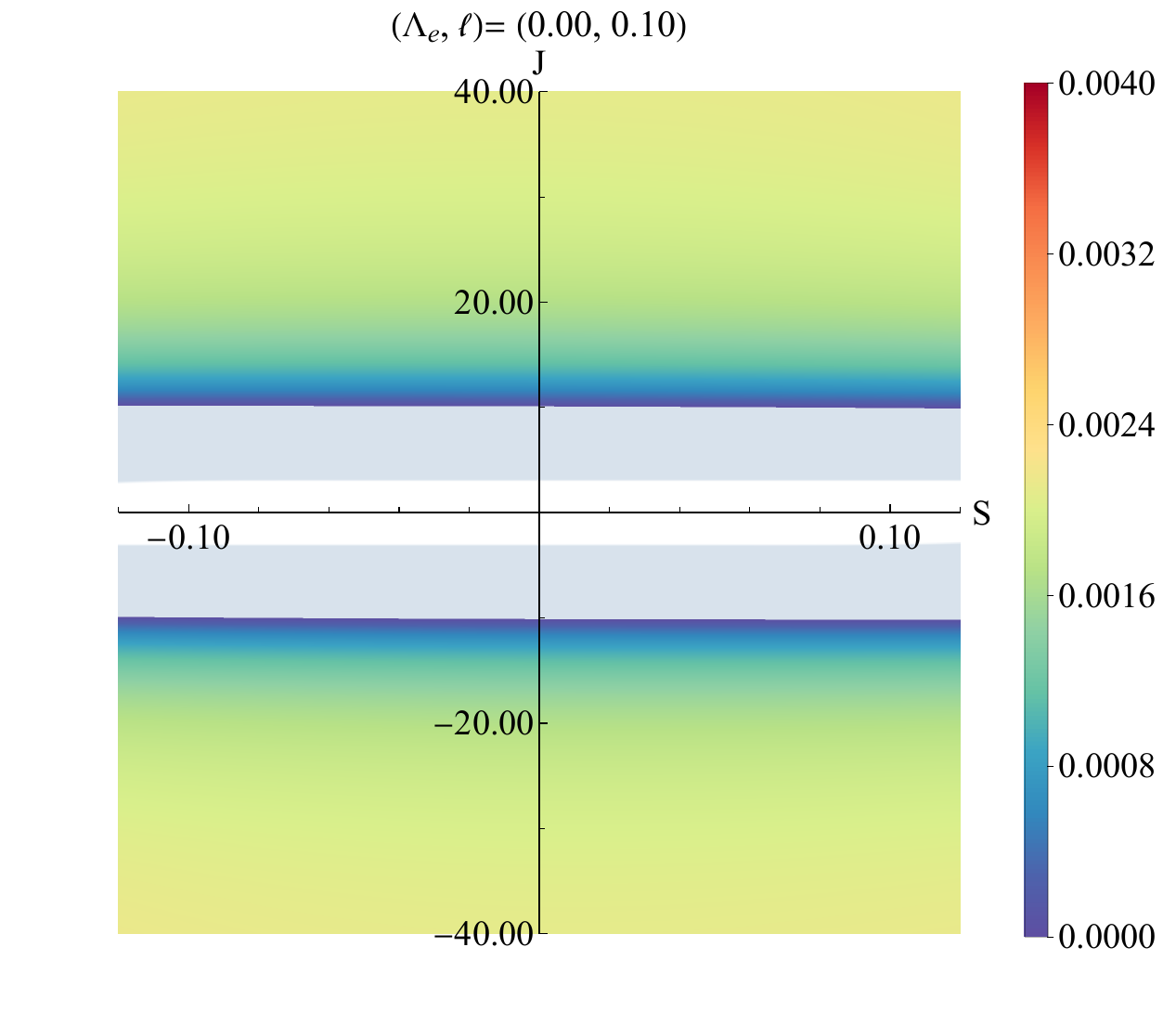}}
	\subfigure[]{\includegraphics[width=0.4\textwidth]{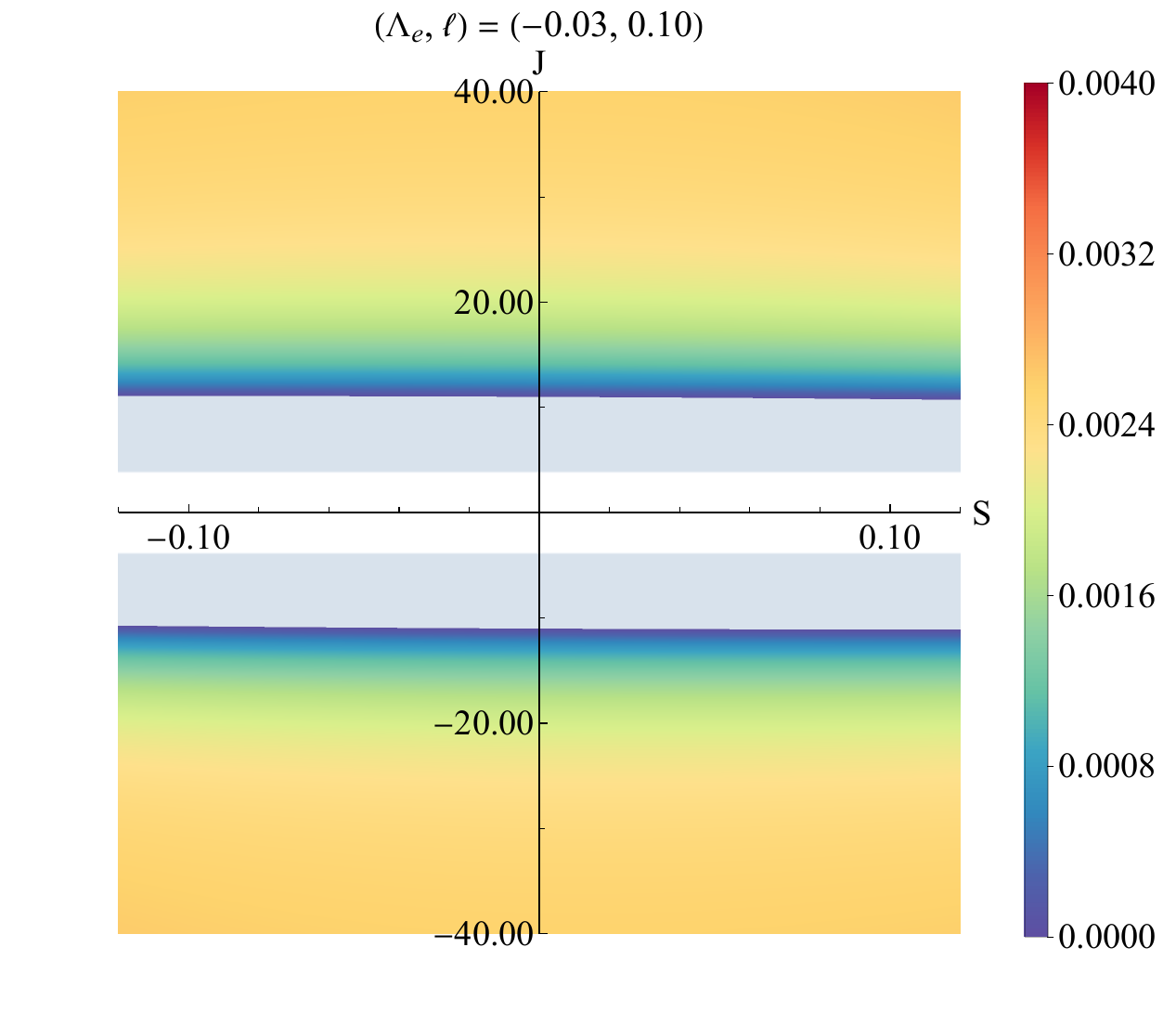}}
	\subfigure[]{\includegraphics[width=0.4\textwidth]{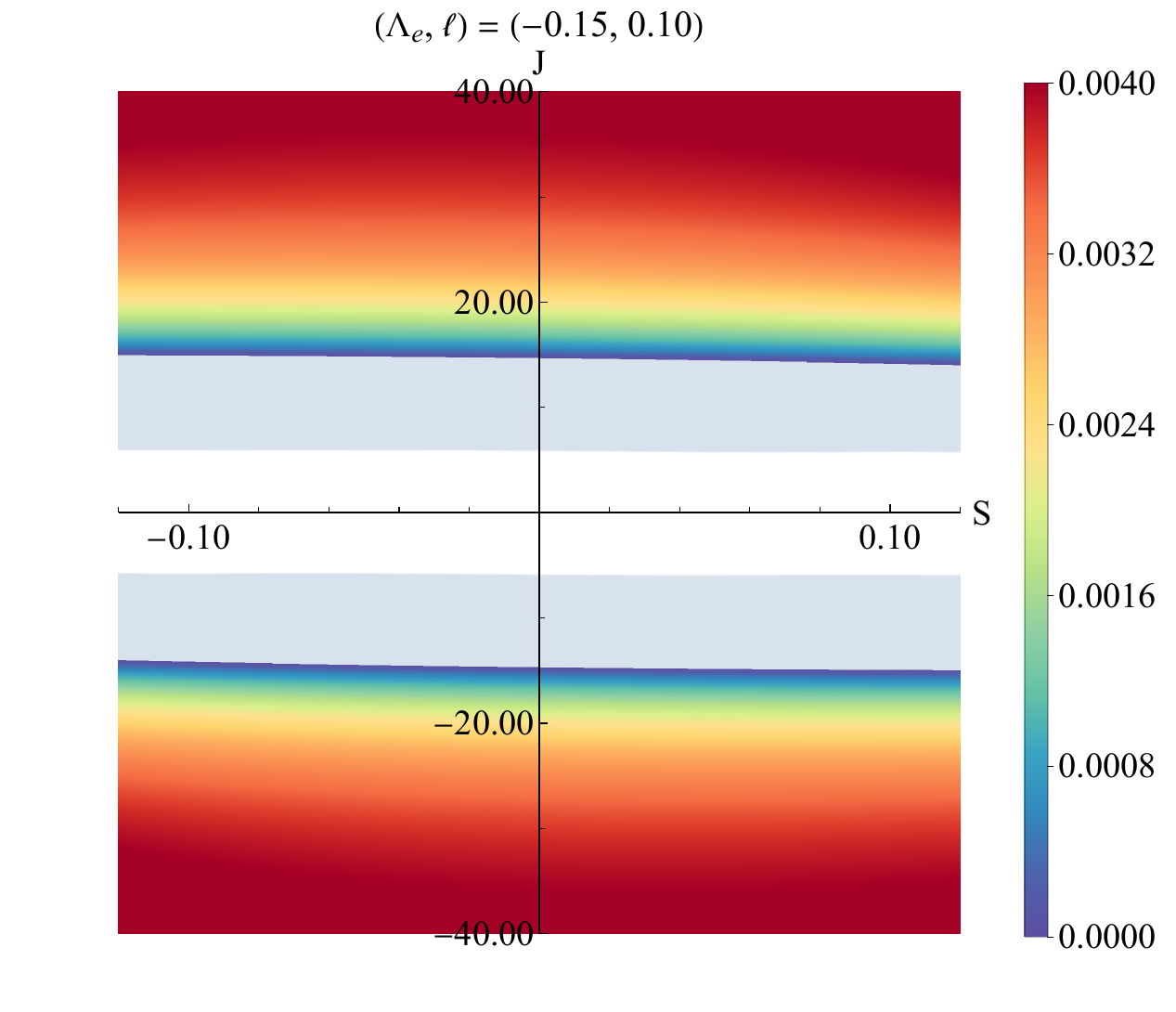}}
\caption{Effects of the particle total angular momentum and spin on the chaos bound violation for $Q_0=0.91$. }
\label{Fig5}
\end{figure*}

\begin{figure*}[t]
	\centering
	
	\subfigure[]{
		\includegraphics[width=0.32\textwidth]{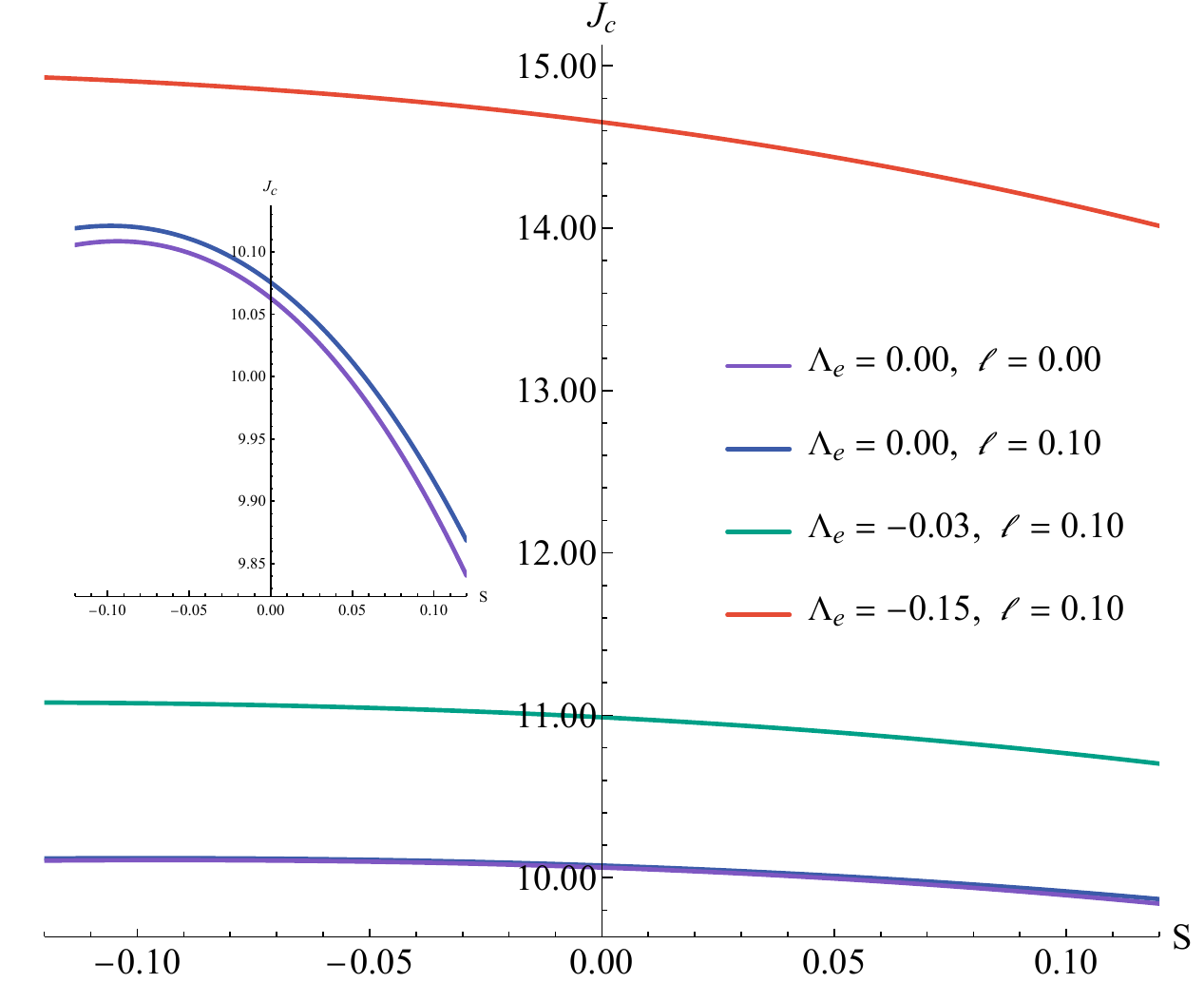}}
		\caption{The critical total angular momentum of the particle as a function of its spin for $Q_0=0.91$.}
	\label{Fig6}
\end{figure*}

\begin{figure*}[t]
	\centering
	
	\subfigure[]{
		\includegraphics[width=0.32\textwidth]{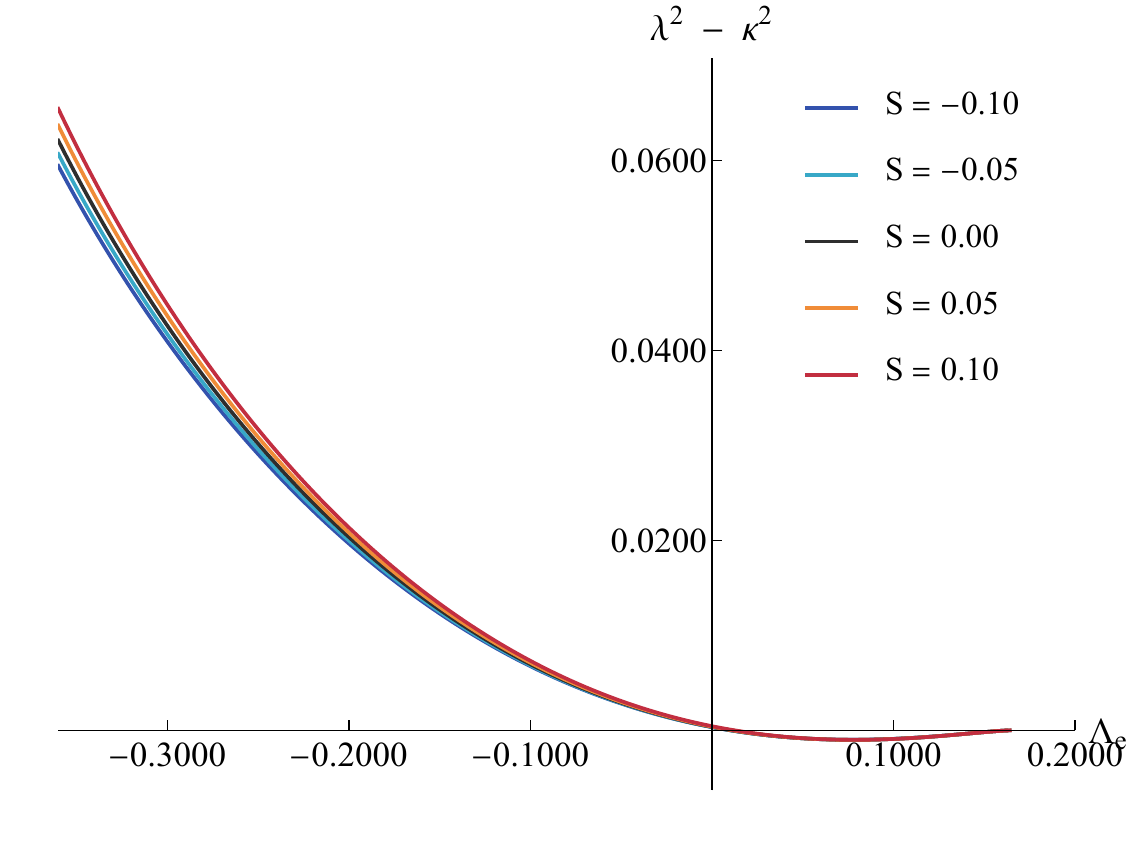}}
		\caption{Effects of the effective cosmological constant on the chaos bound violation for $Q_0=0.94$, $\ell=0.05$ and $J=10.00$.}
	\label{Fig7}
\end{figure*}

For each set of parameters $(Q_0,\ell,\Lambda_e,q,J,S)$, we first solve Eqs.~\eqref{circular1} and~\eqref{circular2} simultaneously for the particle energy $E$ and the circular orbit radius $r_0$. Among the resulting solutions, we retain only the physically admissible unstable circular orbits satisfying $\mathcal{V}_{\mathrm{eff}}''(r_0;E)<0$ and lying in the relevant static region outside the event horizon. The corresponding local orbital LE is then evaluated from Eq.~\eqref{le}, while the surface gravity is calculated from Eq.~\eqref{sf}. By comparing these two quantities, we investigate possible violations of the chaos bound and their dependence on the black hole and particle parameters. Throughout this section, the term “chaos bound violation” refers to the classical local-orbital criterion $\lambda>\kappa$, rather than to a violation of the MSS bound associated with quantum information scrambling. The numerical results are displayed in Figs.~\ref{Fig1}--\ref{Fig7}. Unless otherwise stated, we set $M=1.00$ and $q=0.10$.

Figure~\ref{Fig1} illustrates the combined effects of the black hole charge and the Lorentz-violating parameter on the violation of the chaos bound. The figure contains three panels, where the horizontal and vertical axes denote the Lorentz-violating parameter and the black hole charge, respectively, while the color scale represents the quantity $\Delta=\lambda^{2}-\kappa^{2}$. The gray regions correspond to parameter configurations satisfying the chaos bound, whereas the white regions indicate the absence of unstable equilibrium orbits. This color convention will be used throughout the following analysis. 
As shown in Fig.~\ref{Fig1}(a), the parameter region satisfying the chaos bound is larger than the region where the bound is violated. With increasing Lorentz-violating parameter, both the critical charge associated with the onset of violation, defined by $\Delta=0$, and the extremal charge, corresponding to the maximum allowed black hole charge, decrease simultaneously. Furthermore, the deviation between the exponent and the surface gravity becomes more pronounced as the black hole charge approaches its extremal value. By fixing the particle spin and varying the cosmological constant to $\Lambda_e=-0.03$ and $\Lambda_e=-0.15$, respectively, we obtain Figs.~\ref{Fig1}(b) and \ref{Fig1}(c). These results show that a more negative cosmological constant reduces both the bound-satisfying and bound-violating domains, accompanied by a further decrease in the critical and extremal charges.

The above observations suggest that the Lorentz-violating parameter and the cosmological constant jointly regulate the parameter space in which chaos bound violations can occur. In particular, increasing the Lorentz-violating parameter or enhancing the AdS curvature by taking a more negative cosmological constant reduces the available region for violation, as reflected by the simultaneous reduction of the critical and extremal charges. This behavior can be understood from two aspects. First, Lorentz symmetry breaking modifies the effective spacetime geometry and consequently changes the effective potential governing unstable equilibrium orbits. As a result, the conditions required for violating the chaos bound become more restrictive. Second, a more negative cosmological constant strengthens the confining effect of the AdS background, modifying the structure of the effective potential and narrowing the parameter range that supports unstable equilibrium configurations. Therefore, in Bumblebee gravity, chaos bound violation does not occur universally but depends sensitively on the Lorentz-violating parameter and the asymptotic structure of spacetime. A negative cosmological constant suppresses the parameter region allowing violations by limiting the existence of unstable equilibrium orbits. However, as will be shown below, once the chaos bound is violated, the same AdS curvature can enhance the magnitude of the violation.

Figure~\ref{Fig2} demonstrates the combined effects of the black hole charge and the total angular momentum of the particle on the chaos bound violation. The figure consists of nine panels, where the horizontal and vertical axes represent the black hole charge $Q_0$ and the particle's total angular momentum $J$, respectively. Figure~\ref{Fig2}(a) corresponds to the case of a vanishing cosmological constant and a Lorentz-violating parameter $\ell=-0.10$. It is evident that the parameter domain satisfying the chaos bound is considerably larger than the region where the bound is violated. The violation does not result from an independent increase of either the charge or the angular momentum; instead, it requires a simultaneous adjustment of both quantities within a specific parameter range. The critical charge as a function of the total angular momentum is displayed by the blue curve in Fig.~\ref{Fig3}(a). Keeping the cosmological constant unchanged while increasing the Lorentz-violating parameter to $\ell=0$ and $\ell=0.10$ leads to Figs.~\ref{Fig2}(b) and \ref{Fig2}(c), respectively. The most prominent feature is the significant decrease in the extremal charge, indicating that the maximum allowed charge is strongly affected by the Lorentz-violating parameter. The corresponding critical charge curves are represented by the green and orange curves in Fig.~\ref{Fig3}(a).

Next, fixing $\ell=-0.10$ and increasing the magnitude of the negative cosmological constant to $\Lambda_e=-0.03$ and $\Lambda_e=-0.15$, we obtain Figs.~\ref{Fig2}(d) and \ref{Fig2}(g). In both cases, the boundary separating the bound-violating and bound-satisfying regions moves toward the upper-left direction in the $(Q_0,J)$ parameter space. Meanwhile, the minimum total angular momentum required for the existence of unstable equilibrium orbits increases as the magnitude of the negative cosmological constant becomes larger, whereas the maximum allowed charge decreases. Consequently, the parameter region satisfying the chaos bound becomes progressively reduced. Moreover, the stronger color variation near the maximum charge within the violation region indicates an enhanced deviation between the exponent and the surface gravity. Finally, by fixing the Lorentz-violating parameter at $\ell=0$ and $\ell=0.10$ and varying the cosmological constant, the remaining panels in the second and third columns are obtained. These results exhibit the same qualitative behavior as the first column, confirming the robustness of the dependence of chaos bound violation on the cosmological constant and the Lorentz-violating parameter.

To quantify the boundary shifts of the chaos bound shown in Fig.~\ref{Fig2}, Fig.~\ref{Fig3} presents the critical charge $Q_{0c}$ as a function of the particle's total angular momentum $J$. Figures~\ref{Fig3}(a)--(c) correspond to $\Lambda_e=0$, $-0.03$, and $-0.15$, respectively, with each panel comparing three values of the Lorentz-violating parameter, namely $\ell=-0.10$, $0$, and $0.10$. Figures~\ref{Fig3}(d)--(f) further show the normalized critical charge $Q_{0c}/Q_{\max}$ as a function of the angular momentum, where $Q_{\max}$ denotes the extremal charge determined by the corresponding Lorentz-violating parameter and cosmological constant. Figures~\ref{Fig3}(a)--(c) show that the critical charge decreases monotonically with increasing total angular momentum for all considered values of the cosmological constant. The decrease is more pronounced in the low-angular-momentum regime, while the curves gradually flatten and approach a nearly constant value at higher angular momenta. This behavior indicates a nonlinear dependence of the critical charge on the particle angular momentum. Specifically, increasing the angular momentum substantially lowers the critical charge in the low-to-intermediate angular momentum range, whereas this effect becomes saturated for sufficiently large values. Therefore, the sensitivity of the chaos bound critical boundary to the particle angular momentum is mainly concentrated in the low angular momentum regime. For a fixed cosmological constant, the critical charge curves always satisfy
$Q_{0c}(\ell=-0.10)>Q_{0c}(\ell=0)>Q_{0c}(\ell=0.10)$, indicating that increasing the Lorentz-violating parameter shifts the critical boundary toward smaller black hole charges. In particular, a positive Lorentz-violating parameter reduces the charge required to reach the critical state, whereas a negative parameter shifts the critical point toward larger charges. This ordering remains unchanged for all three cosmological constants considered, demonstrating that the influence of the Lorentz-violating parameter on the critical charge is independent of the specific choice of the cosmological constant. A comparison among Figs.~\ref{Fig3}(a)--(c) further shows that, for a fixed $\ell$, the critical charge curves move downward as the cosmological constant decreases from $0$ to $-0.03$ and $-0.15$. Hence, a stronger AdS curvature effect also reduces the critical charge. This tendency persists over the entire range of angular momentum, suggesting that the cosmological constant mainly changes the overall location of the critical curves rather than their qualitative dependence on the particle angular momentum. Consequently, the Lorentz-violating parameter and the AdS curvature jointly regulate the critical charge: both increasing $\ell$ and making $\Lambda_e$ more negative cause the chaos bound critical state to occur at smaller black hole charges.

Since the extremal charge itself depends on both the Lorentz-violating parameter and the cosmological constant, the absolute value of $Q_{0c}$ alone does not fully characterize the proximity of the critical state to extremality. To clarify this point, Figs.~\ref{Fig3}(d)--(f) investigate the normalized critical charge $Q_{0c}/Q_{\max}$. In the low-angular-momentum regime, the normalized critical charge approaches unity, indicating that the critical state is mainly realized for black holes close to extremality. As the angular momentum increases, the ratio decreases rapidly and gradually approaches a narrow interval around $0.94$ at large angular momentum. This behavior demonstrates that increasing the particle angular momentum not only decreases the absolute value of the critical charge but also reduces the degree of near-extremality required for the onset of chaos bound violation. Therefore, at low angular momentum, the critical behavior is strongly associated with near-extremal black holes, whereas at high angular momentum, the critical state can be achieved in backgrounds relatively far from extremality. Unlike the clear separation among the critical charge curves in Figs.~\ref{Fig3}(a)--(c), the normalized curves in Figs.~\ref{Fig3}(d)--(f) are much closer to each other. This indicates that the main differences in the critical charge caused by different Lorentz-violating parameters originate from the variation of the extremal charge scale. The Lorentz-violating parameter modifies the allowed charge range of the black hole background and consequently shifts the critical charge along the $Q_0$ axis. Although the absolute critical charges differ significantly for different values of $\ell$, their normalized values remain approximately comparable. Nevertheless, the insets reveal a small but systematic separation among the normalized curves, satisfying $(Q_{0c}/Q_{\max})_{\ell=0.10}> (Q_{0c}/Q_{\max})_{\ell=0}> (Q_{0c}/Q_{\max})_{\ell=-0.10},$ which is opposite to the ordering of the unnormalized critical charges. This behavior arises because, although increasing the Lorentz-violating parameter decreases $Q_{0c}$, the corresponding reduction in $Q_{\max}$ is relatively stronger, leading to a slight increase in the normalized ratio. Thus, a positive Lorentz-violating parameter allows the critical state to occur at a smaller absolute charge, while the corresponding critical configuration is relatively closer to extremality when compared with the maximum charge of the same background. Therefore, the near-extremality of the critical state cannot be characterized solely by $Q_{0c}$; instead, the normalized quantity $Q_{0c}/Q_{\max}$ provides a more appropriate measure. Figures~\ref{Fig3}(d)--(f) also demonstrate that the normalized differences are affected by the cosmological constant. For $\Lambda_e=0$, the normalized curves corresponding to different Lorentz-violating parameters nearly overlap, with deviations visible only in the insets. When $\Lambda_e=-0.03$, the separation becomes more apparent, and it is further enhanced for $\Lambda_e=-0.15$. This indicates that a stronger AdS curvature increases the sensitivity of the normalized critical charge to Lorentz symmetry breaking.

Overall, Figs.~\ref{Fig3}(a)--(f) reveal that the variation of the critical black hole charge contains two contributions. The dominant contribution arises from the modification of the extremal charge scale induced by the Lorentz-violating parameter and the cosmological constant, explaining why the curves become nearly coincident after normalization. Meanwhile, the residual separation among the normalized curves indicates that the critical condition also receives additional contributions from the background geometry and the instability properties of particle orbits. Therefore, Fig.~\ref{Fig3} not only provides a quantitative characterization of the dependence of the critical charge on the particle angular momentum, Lorentz-violating parameter, and cosmological constant, but also clarifies that the primary variation originates from changes in the extremal charge scale, while the remaining differences reflect the additional influence of Lorentz symmetry breaking and spacetime curvature on the onset of chaos bound violation.

Figure~\ref{Fig4} illustrates the combined effects of the particle charge and the total angular momentum on the chaos bound violation. Figure~\ref{Fig4}(a) corresponds to the case of a vanishing cosmological constant. It can be seen that the parameter region satisfying the chaos bound is smaller than the region where the bound is violated. As the particle charge varies from negative to positive values, both the gray and white regions gradually shrink. This behavior indicates that the electromagnetic interaction associated with the particle charge modifies the effective potential governing the particle motion, thereby shifting the parameter domain in which unstable equilibrium orbits can exist. Taking the effective cosmological constant as $\Lambda_e=-0.03$ and $\Lambda_e=-0.15$ gives Figs.~\ref{Fig4}(b) and \ref{Fig4}(c), respectively. Similar to the case with vanishing cosmological constant, increasing the particle charge reduces both the gray and white regions. Meanwhile, the color distribution in the chaos bound violating region changes significantly, indicating an enhanced deviation between the exponent and the surface gravity. Furthermore, as the magnitude of the negative cosmological constant increases, the gray region becomes smaller, whereas the white region expands. This suggests that decreasing the AdS curvature radius reduces the parameter space in which the chaos bound is satisfied while increasing the parameter range where unstable equilibrium orbits disappear.

The above results demonstrate that the particle charge and the cosmological constant jointly regulate the chaos bound violation, although their effects arise from different physical origins. The influence of the particle charge is relatively global, compressing both the bound-satisfying region and the region without unstable equilibrium orbits. In contrast, the effect of the cosmological constant is more selective: a more negative cosmological constant mainly suppresses the bound-satisfying region while simultaneously enlarging the domain in which unstable equilibrium orbits are absent. In addition, the enhanced color contrast within the chaos bound violating region indicates that, although a more negative cosmological constant reduces the parameter space allowing violations, it increases the magnitude of $\Delta$ in the remaining violating region. These different behaviors can be attributed to the distinct ways in which the two parameters affect the particle dynamics. The particle charge modifies the motion through electromagnetic interactions, while the cosmological constant changes the spacetime geometry and consequently alters the effective potential structure.

To investigate the combined effects of the particle spin and the total angular momentum on the chaos bound violation, we present Figs.~\ref{Fig5} and \ref{Fig6}. Figure~\ref{Fig5}(a) shows the case with a vanishing cosmological constant, \(\Lambda_e=0\), and \(\ell=0\). It can be seen that the parameter region satisfying the chaos bound is smaller than the region where the bound is violated. As the total angular momentum increases, the deviation between the  exponent and the surface gravity becomes more significant, whereas the influence of the particle spin is relatively weak in this case. The dependence of the critical angular momentum on the particle spin is shown in Fig.~\ref{Fig6}, where the critical angular momentum decreases slowly as \(S\) increases from \(-0.10\) to \(0.10\). Figure~\ref{Fig5}(b) shows the case with a vanishing cosmological constant, \(\Lambda_e=0\), and \(\ell=0.10\). The regions where the chaos bound is satisfied or violated remain qualitatively similar to those in Fig.~\ref{Fig5}(a). However, the color distribution is modified, indicating that the Lorentz-violating parameter \(\ell\) changes the magnitude of \(\Delta\) without substantially altering the overall structure of the bound-satisfying and bound-violating regions. Taking the effective cosmological constant as $\Lambda_e=-0.03$ and $\Lambda_e=-0.15$ gives Figs.~\ref{Fig5}(c) and \ref{Fig5}(d), respectively. Compared with the case of vanishing cosmological constant, the color variation within the chaos bound violating region becomes more pronounced, indicating a further enhancement of the deviation between the exponent and the surface gravity. Meanwhile, the critical total angular momentum required for violating the chaos bound increases, with its dependence on the particle spin displayed in Fig.~\ref{Fig6}. In addition, both the bound-satisfying region and the region without unstable equilibrium orbits become significantly enlarged. To further examine the compatibility of our results with previous studies, we consider the RN limit by setting $\ell=\Lambda_e=0$. For a direct comparison with Ref.~\cite{YCL}, we adopt the same black hole and particle parameters, $Q_0=0.91$ and $q=0.10$. Figure~\ref{Fig6} shows the critical total angular momentum $J_c$ as a function of the particle spin in this limit. In the further spinless limit, $S=0$, we obtain $J_c=10.06$, which agrees with the critical value in~\cite{YCL} for the same RN background and particle parameters. This agreement provides a direct numerical consistency check of our equations of motion and numerical analysis in the spinless RN limit. In the spinless limit, our results can also be compared with the analysis of charged particle circular orbits in RN and RN--AdS black holes presented by Lei and Ge~\cite{LG1}. They found that increasing the particle angular momentum enhances the radial instability of timelike circular orbits and makes the condition $\lambda>\kappa$ more readily satisfied for sufficiently large angular momentum, particularly for near-extremal charged black holes. A qualitatively similar behavior is found in the present analysis for $\ell=0$ and $S=0$. Since Ref.~\cite{LG1} determines the particle charge for a prescribed circular orbit radius, whereas in our analysis the particle charge is fixed and the allowed unstable circular orbit is obtained from the circular orbit conditions, a point-by-point quantitative comparison is not appropriate. Nevertheless, the common dependence on the particle angular momentum and the near-extremal background shows that the two analyses are qualitatively consistent.

These results show that the total angular momentum, the cosmological constant, and the particle spin affect the chaos bound violation in different ways. The total angular momentum primarily determines the strength of the violation: larger values of $J$ lead to a greater deviation of the exponent from the chaos bound. The cosmological constant exhibits a more complicated influence. A more negative cosmological constant increases the critical angular momentum required for the onset of violation and enlarges the parameter domain where unstable equilibrium orbits disappear, thereby reducing the available parameter space for the bound violation. However, for the configurations where the violation persists, the value of $\Delta$ becomes larger, indicating a stronger violation. In contrast, the direct contribution of the particle spin is relatively small. Over the range considered, varying the signed spin parameter produces only a modest shift of the critical condition, with \(J_c\) decreasing gradually as the spin increases from negative to positive values.

Figure~\ref{Fig7} illustrates the dependence of the chaos bound violation on the cosmological constant. It is observed that, in the AdS regime, the quantity $\Delta$ increases with increasing magnitude of the negative cosmological constant. For the representative parameter configuration considered in Fig.~\ref{Fig7}, all calculated LEs remain larger than the surface gravity throughout the AdS region, indicating that the deviation from the chaos bound becomes increasingly significant along the selected parameter trajectory. In contrast, for dS spacetimes, $\Delta$ exhibits a nonmonotonic behavior: it initially decreases and subsequently increases as the cosmological constant increases. When the cosmological constant exceeds a certain critical value, all LEs become smaller than the surface gravity, and the chaos bound is restored. Moreover, as the cosmological constant decreases from positive to negative values, the effect of particle spin on $\Delta$ becomes increasingly pronounced, as reflected by the growing separation among the curves corresponding to different spin values.

The above results demonstrate that the asymptotic structure of spacetime plays an important role in regulating the chaos bound violation, with AdS and dS backgrounds exhibiting qualitatively different behaviors. In AdS spacetime, the negative cosmological constant modifies the effective potential by enhancing the confining effect of the background geometry, thereby increasing the instability of unstable equilibrium orbits once their existence conditions are satisfied. For the specific parameter set adopted in Fig.~\ref{Fig7}, the exponent exceeds the surface gravity throughout the AdS region. However, this behavior should not be interpreted as a universal violation of the chaos bound for all AdS configurations. Instead, it indicates that along the chosen parameter trajectory, the strength of the violation increases with the AdS curvature. For dS spacetime, the presence of the cosmological horizon alters the global spacetime structure and modifies the effective potential governing particle motion. The violation occurs only within a limited range of small cosmological constants. When the cosmological constant becomes sufficiently large, the expansion effect associated with the dS background dominates, the particle trajectories become more stable, and the chaos bound is satisfied. It is also noteworthy that the influence of particle spin depends strongly on the spacetime curvature. In the dS regime, the spin contribution to the chaos bound violation is relatively weak. As the cosmological constant decreases and the background evolves toward the AdS regime, the coupling between the particle spin and spacetime curvature becomes stronger, leading to a more evident separation among the curves with different spin values. Therefore, within the framework of Bumblebee gravity, a more negative cosmological constant reduces the parameter space that permits chaos bound violation by restricting the existence of unstable equilibrium orbits, while simultaneously enhancing the magnitude of the violation in the remaining allowed region. Conversely, a sufficiently large positive cosmological constant restores the chaos bound, and the influence of particle spin becomes increasingly sensitive to the curvature structure during the transition from dS to AdS backgrounds.

\section{Conclusion and discussion} \label{sec4}

In this work, we have investigated chaos bound violations for spinning charged particles in static spherically symmetric black hole spacetimes within Bumblebee gravity. By numerically calculating the LE of spinning particles, we systematically analyzed how the Lorentz-violating parameter, particle spin, and cosmological constant affect the chaotic dynamics. Our results demonstrate that the Lorentz-violating parameter modifies the background geometry, while particle spin introduces an additional modulation mechanism through the spin--curvature coupling, which alters the effective potential and the instability of circular orbits. These effects jointly determine the LE and consequently regulate the occurrence of chaos bound violation.

The Lorentz-violating parameter generally suppresses chaos bound violations. As this parameter increases, the parameter region supporting violation becomes narrower, accompanied by decreases in both the critical charge and the extremal charge. This behavior originates from the modification of the background spacetime geometry induced by Lorentz symmetry breaking, which alters the effective potential governing unstable circular orbits and changes their instability conditions. The normalized analysis further indicates that part of the variation in the critical charge arises from the modification of the extremal charge scale. After normalization, the curves corresponding to different Lorentz-violating parameters become closer, while a residual separation remains, reflecting genuine corrections from the modified geometry. These corrections become more significant for smaller AdS curvature radii, where curvature effects are enhanced. The particle spin plays a fine-tuning role in regulating chaos bound violations. Unlike the total angular momentum, which primarily determines the magnitude of the violation, the direct contribution of particle spin is relatively mild. Its main effect is to shift the critical angular momentum threshold. Over the spin range considered, the critical angular momentum decreases gradually as \(S\) increases from negative to positive values. The influence of spin depends strongly on the background curvature. In the dS spacetime, the spin correction remains relatively weak, whereas in the AdS background the spin--curvature coupling becomes more pronounced, leading to clearer separations among the LE curves for different spin values. The cosmological constant plays a dual role in controlling chaos bound violations. A more negative cosmological constant generally reduces the parameter region where violations occur, while for configurations that still violate the bound, the deviation $\Delta=\lambda^2-\kappa^2$ becomes larger. This indicates that the cosmological constant affects both the occurrence and the strength of chaos bound violations through its modification of the spacetime curvature and the near-orbit dynamics.

Finally, we emphasize that the MSS chaos bound and the LE considered in the present work characterize different physical notions of chaos. The MSS exponent describes the exponential growth of out-of-time-order correlators and quantum information scrambling in thermal quantum systems, whereas the exponent calculated here measures the classical local instability of a spinning charged particle around an unstable circular orbit. Therefore, the condition $\lambda^2>\kappa^2$ found in our analysis should be understood as the local orbital instability rate exceeding the surface gravity scale. It neither implies a breakdown of the MSS quantum chaos bound nor, by itself, establishes global chaotic motion in the full classical phase space. In this sense, our results characterize how Lorentz symmetry breaking and particle spin modify the relation between local orbital instability and the horizon scale.


\end{document}